\documentclass[pdftex,twocolumn,epjc3]{svjour3}          

\RequirePackage[T1]{fontenc}

\smartqed  

\RequirePackage{graphicx}
\RequirePackage{mathptmx}      
\RequirePackage{flushend}
\RequirePackage[numbers,sort&compress]{natbib}
\RequirePackage[colorlinks,citecolor=blue,urlcolor=blue,linkcolor=blue]{hyperref}
\usepackage[caption=false]{subfig}
\usepackage{dcolumn}
\usepackage{bm}
\usepackage{amsmath,amssymb}
\usepackage{widetext}

\journalname{Eur. Phys. J. C}

\newcommand{\sub}[2]{{\text{#1, }#2}}

\newcommand{\theRIGHTwayNLOzero}{\sub{NLO}{0}}
\newcommand{\theRIGHTwayLOone}{\sub{LO}{1}}
\newcommand{\theRIGHTwayNLOone}{\sub{NLO}{1}}

\newcommand{\qprime}{\bm{q}_\bot^{\: \prime}}
\newcommand{\eq}[1]{Eq.~\eqref{#1}}
\newcommand{\tvec}[1]{\mathbf{#1}_\bot}
\newcommand{\ltvec}[1]{#1_\bot}

\newcommand{\ord}[1]{\mathcal{O} \left( #1 \right)}

\newcommand{\infinity}{\infty}
\newcommand{\fig}[1]{Fig.\ (\ref{#1})}
\newcommand{\secref}[1]{Sec.\ \ref{#1}}

\begin{document}

\title{Jet Broadening in the Opacity and Twist Expansions}

\author{Hannah Clayton\thanksref{e1,addr1,addr2}
        \and
        Matthew D.\ Sievert\thanksref{e2,addr3}
        \and
        W.\ A.\ Horowitz\thanksref{e3,addr2}
}

\thankstext{e1}{e-mail: hjc73@cam.ac.uk}
\thankstext{e2}{e-mail: msievert@nmsu.edu}
\thankstext{e3}{e-mail: wa.horowitz@uct.ac.za}

\institute{
        Cavendish Laboratory, University of Cambridge, Cambridge, UK\label{addr1}
        \and
        Department of Physics, University of Cape Town, Rondebosch 7701, South Africa\label{addr2}
        \and
        New Mexico State University, Las Cruces, NM 88011, USA\label{addr3}
}

\date{Received: date / Accepted: date}

\maketitle

\begin{abstract}
We compute the in-medium jet broadening $\langle p_\perp^2\rangle$ to leading order in energy in the opacity expansion.  At leading order in $\alpha_s$ the elastic energy loss gives a jet broadening that grows with $\ln E$.  The next-to-leading order in $\alpha_s$ result is a jet narrowing, due to destructive LPM interference effects, that grows with $\ln^2 E$.  We find that in the opacity expansion the jet broadening asymptotics are---unlike for the mean energy loss---extremely sensitive to the correct treatment of the finite kinematics of the problem; integrating over all emitted gluon transverse momenta leads to a prediction of jet broadening rather than narrowing.  We compare the asymptotics from the opacity expansion to a recent twist-4 derivation of $\langle p_\perp^2\rangle$ and find a qualitative disagreement: the twist-4 derivation predicts a jet broadening rather than a narrowing.  Comparison with current jet measurements cannot distinguish between the broadening or narrowing predictions.  We comment on the origin of the difference between the opacity expansion and twist-4 results.
\end{abstract}

%
\section{Introduction}
%
Hard probes such as jets and leading hadrons have long been promised as critical tomographic tools of nuclear media, both hot and cold, because of their sensitivity to final-state interactions with the medium \cite{Bjorken:1982tu,Gyulassy:2004zy,Wiedemann:2009sh,Majumder:2010qh}.  Enormous experimental progress in measuring hard probes has occurred since the advent of the RHIC era \cite{PHOBOS:2004zne,BRAHMS:2004adc,PHENIX:2004vcz,STAR:2005gfr}, with most spectacularly the observation of a huge suppression of leading light hadrons in central heavy ion collisions that decreases with increasing hadron energy out to the maximum measured $\sim100$ GeV at the LHC \cite{CMS:2016xef,ATLAS:2017rmz,ALICE:2018vuu}. Collaborations have made further enormous progress by investigating the effect of the medium on heavy hadrons \cite{PHENIX:2005nhb,Mustafa:2012jh,CMS:2017qjw,CMS:2017uoy,ALICE:2018lyv,ATLAS:2021xtw}, jet suppression \cite{ALICE:2013dpt,ATLAS:2018gwx,CMS:2019btm}, jet structure \cite{Vitev:2008rz} and sub-structure \cite{Chien:2015hda, Kang:2016ehg} to name just a few.  This wealth of qualitative and quantitatively precise experimental data calls for precise theoretical predictions in order to achieve the goal of making hard probes a precise tomographic tool.

At the same time as this experimental work, significant progress has been made in understanding various theoretical aspects of hard parton propagation in hot and cold media relevant for phenomenological comparison with data.  For interactions at strong coupling, the AdS/CFT correspondence has provided numerous insights \cite{Gubser:2006bz,Herzog:2006gh,Liu:2006ug,Casalderrey-Solana:2006fio,Casalderrey-Solana:2011dxg}.  Assuming that the parton-medium interaction can be described using weak-coupling has led to hundreds, if not thousands, of papers; the problem is extremely complicated with multiple relevant scales. Despite the complicated nature of the weak-coupling approach, a weak-coupling paradigm provides much more theoretical control than a strong-coupling one: the objects of interest are easy to understand, interpret, and manipulate perturbatively.  Given that the scale set by the energy of the leading parton is $\mathcal O(10-100)$ GeV $\gg\Lambda_{QCD}$, and the scale set by the first Matsubara frequency $2\pi T$ is marginal or semi-hard, we expect that  perturbative $\alpha_s\ll1$ methods will describe phenomenological jet energy loss.  Further, energy loss models built on weak-coupling energy loss derivations have seen incredible success in describing a wide range of observables over many orders of magnitude \cite{Horowitz:2012cf,JET:2013cls,Zigic:2021fgf}.  We will thus focus our attention on the weak coupling paradigm in this work.

The usual method of deriving weak-coupling energy loss expressions that are employed in these successful energy loss models \textit{assumes} a trivial factorization that decouples the initial hard production process from the subsequent final-state energy loss processes \cite{Gyulassy:2004zy,Wiedemann:2009sh,Majumder:2010qh,Armesto:2011ht}.  In this picture, subsequent to production, the leading hard parton encounters direct exchanges with the medium degrees of freedom (collisional / elastic energy loss; leading order in $\alpha_s$) and the leading hard parton also suffers from medium-induced radiation (radiative / inelastic energy loss; next-to-leading order in $\alpha_s$).  Asymptotically, the collisional energy loss grows with the logarithm of the leading parton energy.  In the Bethe-Heitler limit, the radiative energy loss grows linearly with energy.  In nuclear collisions the initial hard process that creates the leading high-energy parton involves a rapid and massive acceleration of an object charged under SU(3); the hard production process necessarily generates a huge amount of so-called vacuum radiation.  The subsequent medium-induced radiation quantum-mechanically destructively interferes with this vacuum radiation; the growth of the radiative energy loss is softened from linear in $E$ to logarithmic, which is known as the Landau-Pomeranchuk-Migdal (LPM) effect.  Therefore both elastic and inelastic energy loss are equally important at large energies, with the radiative energy loss about 4 times larger than the collisional in phenomenologically relevant physical situations \cite{Horowitz:2010dm}.

The above qualitative estimates for the growth in energy of the energy loss are reproduced quantitatively by many different derivations of both collisional and radiative energy loss within this energy loss picture that assumes a factorization of the production process from the final-state interactions with the medium.  A proper subset of the derivations of elastic energy loss include \cite{Bjorken:1982tu,Braaten:1991we,Thoma:1990fm}.  A proper subset of the inelastic energy loss derivations includes \cite{Baier:1996sk,Zakharov:1997uu,Gyulassy:2000er,Wiedemann:2000za,Arnold:2000dr,Wang:2001ifa}.  Reviews include \cite{Wiedemann:2009sh,Majumder:2010qh}.  In this work, we will focus on the opacity expansion picture for computing the collisional and radiative processes affecting leading parton propagation.  Roughly speaking, the opacity expansion is an expansion in the number of interactions $L/\lambda$, where $\lambda$ is the gluon mean free path, a hard parton has with the soft in-medium quasiparticles \cite{Gyulassy:2000er,Djordjevic:2003zk}.  We choose to focus on the opacity expansion because it naturally incorporates the LPM effect, takes into account the finite kinematics in phenomenologically relevant processes, and has a relatively simple closed form for the radiated single inclusive gluon distribution at first order in opacity.

For many years the field has sought to build upon these qualitative comparisons between energy loss models and data to achieve quantitative comparisons \cite{JET:2013cls,JETSCAPE:2017eso,JETSCAPE:2021ehl}.  One avenue of research has attempted to quantify the various systematic theoretical uncertainties associated with currently used energy loss derivations \cite{Aurenche:2008hm,Horowitz:2009eb,Armesto:2011ht}.  Less attention has been paid to critically examining the basic assumptions made in deriving the energy loss formulae.  Further, there are highly non-trivial unresolved conceptual issues related to placing energy loss derivations on more rigorous footing.  What one would really like is a systematic order-by-order expansion of specific hard probe observables in some small quantity.  Generally speaking, the paradigm one has in mind for these hard probes of media is that of collinear factorization.  

On the other hand, we consider the potential application of the collinear factorization framework to jet observables in nuclear processes.  Collinear factorization is a highly developed field that is central to ep and eA phenomenology \cite{Collins:2011zzd}.  The strength of this field rests on factorization theorems.  A factorization theorem proves to all orders in $\alpha_s$ that one may expand an observable in inverse powers of a large scale $Q$.  In collinear factorization, there is a convolution of a short-distance hard cross section with long-distance, non-perturbative objects such as parton distribution functions and/or fragmentation functions.  Crucially, the hard cross sections are perturbatively computable order-by-order in $\alpha_s$.  And while the non-perturbative objects cannot be computed themselves from first principles, their evolution equations in $Q$ are computable order-by-order in $\alpha_s$.  The essential ingredient in the proof of collinear factorization is the cancellation of (nonperturbative) soft gluon radiation which entangles different sectors of the scattering process.  This cancellation of soft gluon entangling radiation ``quarantines'' non-perturbative QCD physics into a small number of \emph{universal} long-distance objects.  

The expansion in powers of $\ell_\perp/Q$, where $\ell_\perp\sim\Lambda_{QCD}$ is some typical transverse momentum scale in the problem, is known as the twist expansion.  For example a twist-2 calculation may receive corrections only up to $\mathcal O\big((\ell_\perp/Q)^3\big)$.  Factorization has been rigorously proven for several observables at leading twist (twist-2, or $\ord{\left(\Lambda_{QCD}/Q\right)^0}$) and for various spin asymmetries at twist-3, or $\ord{\left(\Lambda_{QCD}/Q\right)^1}$ \cite{Collins:1989gx, Qiu:1991pp, Qiu:1991wg, Collins:1996fb, Collins:1998be, Collins:2011zzd}.  

So far, there has been no rigorous factorization-like proof for any medium modified hard probe observable.  I.e.\ the assumption of a factorization of the hard production process from the subsequence in-medium propagation is currently uncontrolled.  Further, without an overarching theoretical framework such as collinear factorization, it is difficult to know how to expand order-by-order in the various competing energy loss expansion parameters such as $\alpha_s$ or the opacity $L/\lambda$.  Should this factorization of hard production and subsequent evolution be valid, there should be a corresponding factorization theorem.  One would hope that a factorization theorem in energy loss would provide just the necessary framework for a well controlled expansion for various energy loss observables as well as a set of \emph{universal} quantities valid across a variety of processes.  Some work has attempted to incorporate ideas from collinear factorization into energy loss-type calculations: e.g.\ the assumption of a factorization of the production process is kept, while the final state energy loss is incorporated into medium modified fragmentation functions that are evolved using DGLAP evolution, often with medium modified splitting functions \cite{Guo:2000nz,Wang:2001ifa,Armesto:2007dt,Aurenche:2008hm,Majumder:2009ge,Kang:2014xsa,Chien:2015vja, Zhang:2018kkn, Zhang:2019toi,Sirimanna:2021sqx}. 

\begin{figure}
    \centering
    \includegraphics[width=\columnwidth]{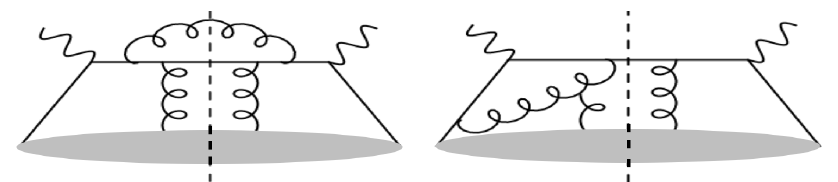}
    \caption{Two of the contributing Feynman diagrams to the transverse momentum broadening in SIDIS at twist-4 (reproduced from \cite{Kang:2013raa}). The left diagram is topologically equivalent to the one used for energy loss in Figure~\ref{fig:diagram}. The right diagram includes cross talk between the parton production and subsequent evolution, which is neglected in energy loss formalisms.}\label{fig:ZB}
\end{figure}

On the other hand, a twist-4 collinear factorization derivation was performed for the jet momentum broadening in semi-inclusive deep inelastic scattering (SIDIS) \cite{Kang:2013raa,Xing:2014kpa,Kang:2014ela}.  \linebreak While this derivation didn't rigorously prove a factorization theorem, the calculation did see that all IR and UV divergences were safely absorbed at next-to-leading order in $\alpha_s$.  Surprisingly, in this calculation the first nontrivial rescattering in the medium is expressed through the appearance of a ``double PDF,'' the four parton correlator containing both the partons participating in the hard scattering and the gluons participating in the rescattering \cite{Kang:2014ela,Xing:2014kpa,Kang:2013raa}.  Critically, this work derived the evolution equation for the four-parton correlator and found a structure significantly different from the standard DGLAP evolution equations which characterize \linebreak twist-2 collinear PDFs, fragmentation functions, and jet functions.  Taken at face value, this result offers a significant challenge to the usual energy loss derivations.  If one merely medium modifies twist-2 fragmentation functions, then one misses entirely the four parton correlators that naturally \linebreak emerge in the twist-4 framework.  Moreover, in the twist-4 derivation there are diagrams that explicitly mix the initial and final state processes, thereby violating a naive assumption of factorization as assumed in energy loss calculations; see \fig{fig:ZB}.  

The twist expansion in collinear factorization appears to correspond in some limit to the opacity expansion.  The hard scattering itself is leading twist (twist-2); the first final-state rescattering (first order in opacity) is twist-4; and successive multiple scatterings (higher orders in opacity) are correspondingly further suppressed in the twist expansion.  One goal of this work is to see if one can make the matching of the twist-4 collinear expansion to the first-order-opacity energy loss calculation more explicit and/or rigorous.

Assuming that the twist-4 approach is ``more correct'' than the usual energy loss approach since the assumption of a factorization of the initial and final state processes is not made \textit{a priori}\footnote{On the other hand, the opacity expansion of the energy loss approach fully captures the LPM effect, which corresponds to a resummation of diagrams in the twist expansion approach.  One could argue that fully capturing the LPM effect via the opacity expansion is a ``more correct'' foundation to build on \cite{Qiu:2003vd}.}, one would like to assess the importance of these terms not present in the energy loss derivation; in particular, perhaps these terms may be small and readily neglected, providing further support to the usual energy loss approach.  In order to make contact with the twist-4 approach from the energy loss approach, we compute the same jet momentum broadening observable from within the opacity expansion.  We find that the two approaches agree exactly at leading order in $\alpha_s$, i.e.\ when one considers only collisional energy loss.  However, we find that the two approaches qualitatively differ at next-to-leading order.  While the growth in energy is the same, the twist-4 approach qualitatively differs from the energy loss approach in that 1) the twist-4 approach includes color triviality breaking terms not present in the energy loss calculation and 2) even when neglecting the color triviality breaking terms in the twist-4 approach, the twist-4 approach yields a jet broadening, whereas the energy loss calculation predicts jet narrowing.  

Interestingly, the prediction of jet transverse momentum narrowing from the energy loss picture is delicate to tease out of the analytic expressions.  We show that if one too carelessly makes the usual assumption that $k_{\perp,\,\mathrm{max}}\sim x\,E$ can be taken to $\infinity$,  then one gets a prediction of jet broadening rather than jet narrowing.  Thus the qualitative prediction from the energy loss calculation is very sensitive to the treatment of the finite kinematic limits for the radiated gluon transverse momentum.  Since the twist-4 approach appears to assume \textit{a priori} that one may safely integrate over all $k_\perp$ up to infinity, we speculate that the twist-4 prediction of jet broadening may be an artifact of this infinite kinematics assumption.

Specifically, after the careful treatment of finite kinematics, we find that the opacity expansion predicts a jet narrowing due to radiative corrections as 
\begin{align}
    \label{eq:ll}
    \langle p_\perp^2\rangle\sim-C_R\alpha_s\frac L\lambda \mu^2 \ln^2\big(\frac{E}{\mu^2 L}\big).
\end{align}
Superficially, taking $\mu^2/\lambda = \hat q$, \eq{eq:ll} appears similar to work that found radiative corrections led to a double logarithmic enhancement to jet broadening \cite{Liou:2013qya} or to work which reabsorbed a double logarithmic enhancement to jet broadening into the jet transport coefficient $\hat q$ \cite{Blaizot:2014bha,Iancu:2014kga,Blaizot:2019muz}.  The most important difference between these works and ours is that none of these works include the physics of the initial hard scattering (and subsequent emission of vacuum-like radiation).  Rather, these calculations assume the existence of a high-energy parton for all time, but which enters a slab of nuclear material at a finite time.  In contrast, in our work we explicitly include the vacuum production radiation necessary for a comparison to hadronic collision measurements.  To further drive home the point, \cite{Liou:2013qya,Blaizot:2014bha,Iancu:2014kga,Blaizot:2019muz} all predict jet broadening, in contradistinction to our finding of jet narrowing.  When the finite creation time and the full kinematics are taken into account, the light-cone path integral formalism of BDMPS-Z also predicts jet narrowing in a dense medium \cite{Zakharov:2018rst,Zakharov:2019fov,Zakharov:2020sfx}.  Second, only \cite{Blaizot:2019muz} considers the case of a few hard scatterings, although, again, without the associated initial state radiation; the others only consider the dense medium saturation limit.  Given that estimates of the mean free path in even the hottest LHC fireballs are $\mathcal O(1)$ fm, phenomenologically relevant calculations are likely closer to the dilute rather than dense limit.  Possibly worse, the dense limit calculations make the harmonic oscillator approximation, which completely misses the power law tails associated with perturbative scattering.

Third, the arguments of the double logarithms found here and in \cite{Liou:2013qya,Blaizot:2019muz} for $\langle \Delta p_T^2\rangle$ are completely different with completely different physical interpretations.  For us, the argument of the double logarithm is $E/\mu^2 L$.  Importantly, the $\mu$ and $L$ dependencies of this argument do not come from kinematic limits.  One is tempted to interpret the argument as a double logarithmic enhancement of the transport coefficient $\hat q$ by the ratio of the gluon formation time $\tau_{form}=xE/k_T^2\sim E/\mu^2$ for moderate $x$ to the length of the medium $L$.\footnote{One should distinguish between the typical formation time for the soft gluons of $x\sim \mu/E$, $\tau_{typ}\sim1/\mu$, that do not affect the transverse broadening much as compared to the more rare, harder emissions with $x\sim\mathcal O(1)$ that are argued to affect the jet transverse broadening \protect\cite{Blaizot:2019muz}.}  On the other hand, \cite{Liou:2013qya} find an argument of the double logarithm due explicitly to phase space limitations of $L/\ell_0$ where $\ell_0$ is a minimum propagation distance and is given by the nucleon size in cold nuclear matter and by the inverse temperature in hot nuclear matter.  Thus the argument of their double logarithm is composed purely of medium properties and is energy independent.  In \cite{Blaizot:2019muz}, the authors claim that in the dilute limit (e.g.\ in an opacity calculation), the argument of one of the double logarithms depends only on medium properties while the argument of the second logarithm is $\mathbf p^2/\mu^2$, where $\mathbf p$ is the measured transverse momentum of the hard particle.  

We attempt to distinguish between the two very different qualitative predictions for $\langle \Delta p_\perp^2\rangle$ from the twist-4 approach and the opacity expansion by comparing to recent ALICE data on jet momentum broadening \cite{ALICE:2015mdb}.  At the present stage, the uncertainties in the ALICE data yield a result consistent with both broadening or narrowing of jet transverse momenta.

%
%

%
\begin{figure}
    \centering
    \includegraphics[width=0.7\columnwidth]{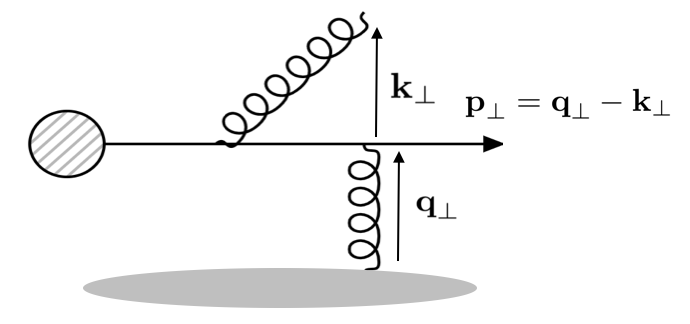}
    \caption{After the initial hard process, the deconfined parton traverses the medium. Transverse momentum can be transferred to this parton via collisions with the medium ($\textbf{q}_{\perp}$) or the stimulated emission of soft gluons ($\textbf{k}_{\perp}$).}\label{fig:diagram}
\end{figure}
%

The rest of this paper is organized as follows.  In Sec.~\ref{sec:model} we summarize the DGLV model and its application to energy loss and momentum broadening at first order in opacity.  In Sec.~\ref{sec:numerics} we evaluate these expressions numerically, keeping the full kinematic bounds on the integrals with no further approximation.  Results are given for the collisional and radiative contributions to the broadening, including the surprising feature that the coefficient appearing in Eq.~\eqref{e:guess} is negative (medium-induced narrowing, rather than broadening).  In Sec.~\ref{sec:analytics} we compute the leading high-energy asymptotics within the DGLV model analytically, illustrating explicitly how sensitive the result is to different choices of the kinematic limits of integration, and carefully reproducing the numerical results of Sec.~\ref{sec:numerics}.  In Sec.~\ref{sec:discussion} we compare the results obtained here with the twist-4 formalism of Refs.~\cite{Kang:2014ela, Kang:2013raa,Xing:2014kpa} and with experimental data.  Finally, we conclude in Sec.~\ref{sec:concl}.

%
\section{The Opacity Expansion}
\label{sec:model}
%

We consider two ways in which jet broadening can occur from an energy loss perspective. First, a parton propagating through the medium can undergo elastic scattering off the medium constituents, broadening the jet transverse momentum distribution through direct exchange with the medium.  We refer to this as the ``collisional'' or ``leading order (LO)'' momentum broadening. Second, interactions with the me\-di\-um can stimulate the emission of gluons off the jet parton, broadening the jet transverse momentum distribution through the recoil against the emitted radiation.  This broadening is referred to as ``radiative'' or ``next-to-leading order (NLO)''.  The contribution to jet broadening arising from the interactions with the medium must be carefully distinguished from the radiative broadening (as in Sudakov emissions) which can occur even in vacuum.

In the opacity expansion, the leading parton or jet process is expanded in numbers of interactions with the medium.  To wit, at zeroth order in opacity, the jet amplitude has no interactions with the medium; we'll denote zeroth order in opacity with a subscript ``0.''  At first order in opacity, the jet process contains \textit{two} interactions with the medium.  These two interactions could come from one interaction in the amplitude and one in the complex conjugate amplitude.  These two interactions could also occur in the amplitude with none in the conjugate amplitude, or vice-versa.  We'll denote first order in opacity with a subscript ``1.''

\subsection{Collisional (LO) Momentum Broadening}
The defining feature of collisional, or ``leading order'' (LO), momentum broadening for us is the lack of any radiation in the process.  Thus the $\alpha_s$ power counting goes as $\alpha_s^n$ for the $n^\mathrm{th}$ order in opacity contribution to the collisional momentum broadening, even though we refer to these processes as ``leading order.''

\subsubsection{\texorpdfstring{0$^{th}$}{0th} Order in Opacity}
\begin{figure}
    \centering
    \includegraphics[width=0.7\columnwidth]{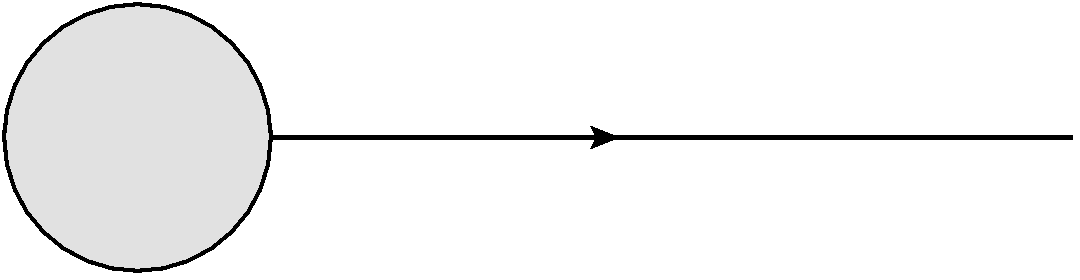}
    \caption{\label{fig:collisionalzeroth} The zeroth order in opacity contribution to the amplitude for leading order momentum broadening.}
\end{figure}

\fig{fig:collisionalzeroth} shows the diagram contributing to collisional broadening at zeroth order in opacity.  Notice that in the energy loss approach the production process for the high-energy parton is represented by a blob, and there is no communication between the blob and the subsequent evolution of the parton.  Since we are interested in the modification of the jet due to the presence of the medium, the zeroth order in opacity contribution to leading order jet broadening is trivial,
\begin{align}
    \langle \ltvec p^2 \rangle_{\sub{LO}{0}} = 0.
\end{align}

%
%

\subsubsection{\texorpdfstring{1$^{st}$}{1st} Order in Opacity}\label{sec:LO1st}
\begin{figure}
    \centering
    \includegraphics[width=0.7\columnwidth]{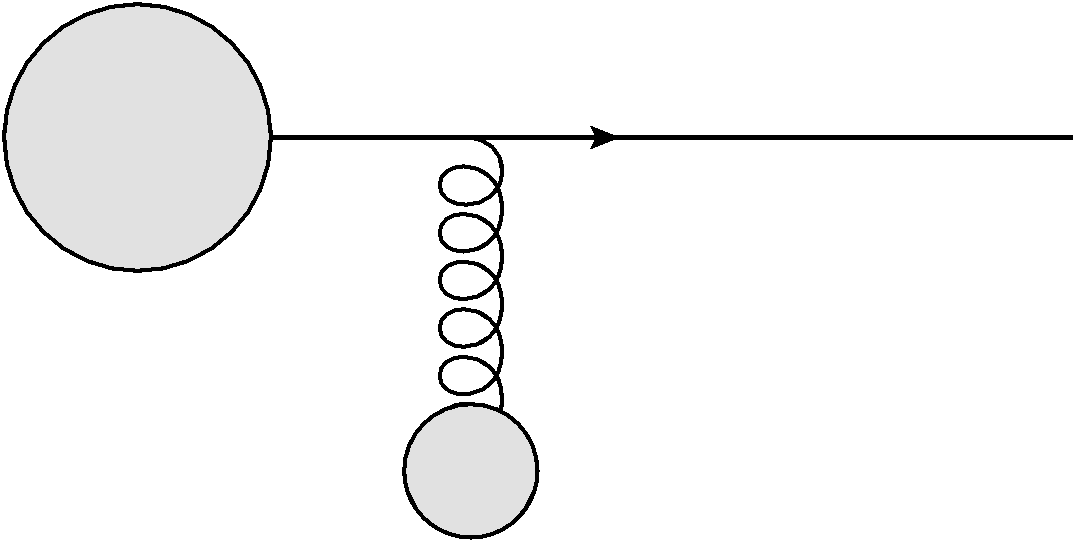}
    \caption{\label{fig:collisionalfirst} The first order in opacity contribution to the amplitude for leading order momentum broadening.}
\end{figure}

\fig{fig:collisionalfirst} shows the diagram contributing to collisional broadening at first order in opacity.  In principle, one must also include the zeroth order in opacity diagram shown in \fig{fig:collisionalzeroth} along with the diagrams with two interactions with the medium (often referred to as ``double Born'' diagrams); when the amplitude is squared, the diagrams with zero and two interactions interfere and contribute to the same order in opacity as the diagram shown in \fig{fig:collisionalfirst}.  However, since the leading order zeroth order in opacity amplitude leaves the hard parton unchanged, the double Born diagrams can only interfere with the zeroth order diagram if they also leave the hard parton unchanged.  

Consistent with other opacity expansion derivations \cite{Gyulassy:2000er,Djordjevic:2003zk} we model the interaction with the medium shown in \fig{fig:collisionalfirst} as a Gyulassy-Wang static scattering center \cite{Gyulassy:1993hr}.  The elastic differential cross section to first order in opacity is then
\begin{align}   \label{e:GW}
    \left.\frac{d^2\sigma^{qg\rightarrow qg}}{ d^2\textbf{q}_\perp}\right|_{1} = \frac{2\alpha_s^2}{(\textbf{q}_\perp^2+\mu^2)^2},
\end{align}
where $\mu\approx gT$ is the chromoelectric Debye screening mass of the medium and $\tvec q$ is the transverse momentum of the $t$-channel gluon exchanged with the medium \cite{Djordjevic:2003zk}; see \fig{fig:diagram} for a schematic of the momenta used in our formulae.  Note that we have assumed that the medium contains only gluons, although this assumption can be straightforwardly relaxed.  

On average, the leading order momentum broadening is given by the mean transverse momentum squared picked up per elastic collision times the number of collisions.  The mean momentum squared picked up \textit{per collision} is given by the weighted average
\begin{align}
    \nonumber
    \int d^2\tvec q  \, \tvec q^2 \frac{d^2\sigma^{qg\rightarrow qg}}{ d^2\tvec q}\bigg/\int d^2\tvec q  \,  \frac{d^2\sigma^{qg\rightarrow qg}}{ d^2\tvec q}.
\end{align}
The number of elastic collisions suffered by the hard parton is given by $L/\lambda$, where $L$ is the length of the medium and $\lambda$ is the mean free path of the hard parton.  Using the first order in opacity result for the leading order cross section, the leading order, first order in opacity transverse momentum jet broadening is
\begin{align}
    \label{eqn:pT2LO}
    \langle \ltvec p^2 \rangle_{\sub{LO}{1}} \equiv \frac{L}{\lambda} \int d^2\tvec q  \, \tvec q^2 \frac{d^2\sigma^{qg\rightarrow qg}}{ d^2\tvec q}\bigg/\int d^2\tvec q  \,  \frac{d^2\sigma^{qg\rightarrow qg}}{ d^2\tvec q}.
\end{align}

Imposing the kinematic limit
\begin{align}
    \label{eq:qmax}
    q_{\text{max}}&=\sqrt{3E\mu}\simeq\sqrt{6ET},
\end{align}
which comes from the maximum $t$ channel exchange for an incoming particle of momentum $E$ and of another particle of momentum $3\mu\sim6T$, from $\mu\sim g\,T$ and $g(2\pi T)\approx2$ for $T\sim400$ MeV, one finds asymptotically
\begin{align}   \label{e:CollAsym}
    \langle p_T^2 \rangle_{\text{LO, }1} 
    &= \frac{L\mu^2\big(3E+\mu\big)\big(\ln(\frac{3E+\mu}{\mu}) - \frac{3E}{3E+\mu}\big)}{3 E\lambda} 
    \notag \\ &\approx
    \frac{L\mu^2}{\lambda} \ln(\frac{E}{\mu}) 
    \notag \\ &=
    \hat{q} L,
\end{align}
where $\hat{q} = \frac{\mu^2}{\lambda} \ln\frac{E}{\mu}$ is the momentum broadening of a quark per unit path length.  Since $\hat q$ denotes the rescattering in the gluon field of the in-medium scattering center which generates the elastic cross section \eqref{e:GW}, it is natural that $\hat{q}$ can be expressed in terms of the gluon PDF \cite{Baier:2002tc}.  For the Gyulassy-Wang model one assumes that the target is composed of heavy, static partons whose gluon distribution can be readily calculated in pQCD\footnote{This expression is derived in the leading logarithmic approximation at small $x_g$, where $x_g$ denotes the momentum fraction of the \textit{gluons} being exchanged with the target, which is in general $x_g \sim \tfrac{\mu}{E}$ even if the Bjorken variable $x_B$ is large $x_B \sim \ord{1}$.} \cite{Kovchegov:2012mbw} to be
\begin{align}
    xG = \frac{2 \alpha_s C_R^\prime}{\pi} \ln\frac{E}{\mu} \: ,
\end{align}
where $C_R^\prime$ is the color Casimir of the medium partons.  The momentum broadening per unit length $\hat{q}$ can then be written in terms of the gluon distribution as
\begin{align}
    \hat{q} = \mu^2 \frac{L}{\lambda} \frac{\pi}{2 \alpha_s C_R^\prime} \, x G \: . 
\end{align}
Thus we see that the large logarithm arising in the collisional broadening term \eqref{e:CollAsym} is associated with the large number of gluons produced by a single parton $\sim dk_\bot^2 / k_\bot^2$.  Physically, the mean transverse momentum squared \textit{per gluon} is of order $\sim \ord{\mu^2}$, the number of gluons $x G$ \textit{per scattering center} is of order $\alpha_s \ln\tfrac{E}{\mu}$, and the typical number of \textit{scatterings in the medium} is $\sim \tfrac{L}{\lambda}$.  Together this reasoning gives $\langle p_T^2 \rangle_\theRIGHTwayLOone \sim \mu^2 \ln\frac{E}{\mu} \tfrac{L}{\lambda}$ as in \eq{e:CollAsym}.  

We'll truncate our opacity expansion at first order for two reasons.  First, most radiative energy loss calculations truncate at this order, and, second, truncation at this order allows one to best make contact with the twist-4 calculation.

%
%

%
\subsection{Radiative (NLO) Momentum Broadening}
%
The defining feature of radiative processes is the emission of one or more gluons; i.e.\ in addition to the original hard parton, there are one or more final state gluons.  In general the goal of the opacity expansion approach is to compute the differential distribution of the radiated gluons.  The production process is assumed factorized and unaffected by the emission of the (predominantly) soft and collinear gluon radiation; the final state of the hard parton is integrated over.  As a result, the derived gluon distribution is inclusive.  Therefore, as will be seen below, the predicted average number of emitted gluons is not fixed.  

%
\subsubsection{Vacuum Emissions (\texorpdfstring{0$^{th}$}{0th} Order in Opacity)}
%

%
\begin{figure}
    \centering
    \includegraphics[width=0.7\columnwidth]{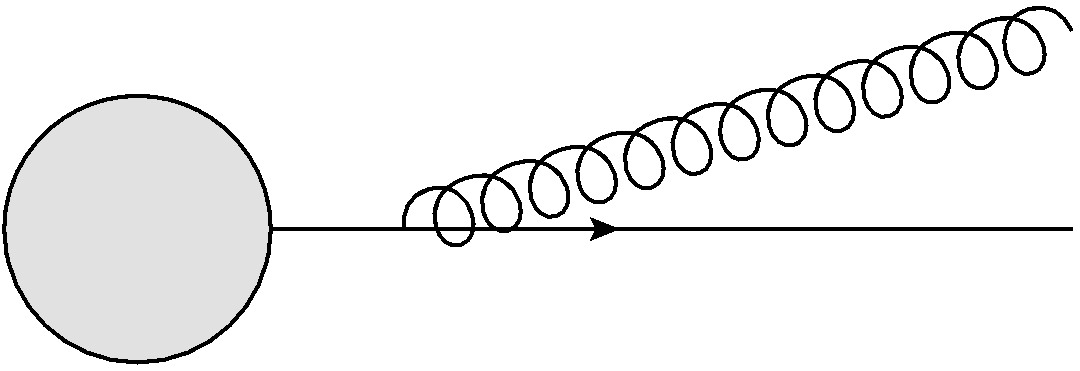}
    \caption{\label{fig:radiativezeroth} The zeroth order in opacity contribution to the amplitude for next-to-leading order momentum broadening.}
\end{figure}
%
\fig{fig:radiativezeroth} shows the diagram contributing to radiative broadening at zeroth order in opacity.  %
The distribution of gluons with energy fraction $x \ll 1$ and transverse momentum $\tvec{k}$ emitted by such a high-energy parton in vacuum is given by
\begin{equation}    \label{eqn:dNvac}
\begin{split}
&\frac{d^3N_g^{(0)}}{dx d^2\textbf{k}_\perp} 
= \frac{C_R \alpha_s}{\pi^2 x} \frac{\textbf{k}_\perp^2}{(\textbf{k}_\perp^2+m_g^2+M^2x^2)^2} \, ,
\end{split}
\end{equation}
where $C_R$ is the quadratic Casimir factor in the color representation $R$ of the jet parton, $M$ is the mass of the jet parton (potentially a heavy quark), and $m_g$ is a mass associated with the radiated gluon arising from the Ter-Mikayelian effect \cite{TerMik:1954,TerMik:1972,Djordjevic:2003be}.

The elementary splitting function \eqref{eqn:dNvac} dictates both the radiative broadening and energy loss in vacuum.  The integral of the distribution \eqref{eqn:dNvac} just gives the average number of radiated gluons $\langle N_g \rangle_0$.  The fraction $\left. \Delta E / E \right|_{\text{NLO, }0}$ of the initial jet energy $E$ which is carried away by the gluon radiation is similarly obtained by computing the average energy fraction $x$ from the distribution \eqref{eqn:dNvac}.  Likewise the mean-square transverse momentum broadening $\langle p_T \rangle_{\text{NLO, }0}$ produced by the emissions is obtained from \eqref{eqn:dNvac} by computing the mean-square momentum $k_\bot^2$ of the radiated gluons:
\begin{subequations}    \label{e:vackern}
\begin{align}
    \langle N_g \rangle_0 &\equiv \int dx \, d^2 \tvec{k}  \:
    \frac{d^3 N_g^{(0)}}{dx \, d^2\textbf{k}_\perp} \: ,
    \\
    \left. \frac{\Delta E}{E} \right|_{\text{NLO, } 0} &=
    \int dx \, d^2 \tvec{k} \: \left( x \, \frac{d^3N_g^{(0)}}{dx \, d^2\textbf{k}_\perp} \right) \: ,
    \\
    \langle p_T^2 \rangle_{\text{NLO, }0} &=
    \int dx \, d^2 \tvec{k} \: \left( k_\bot^2 \, \frac{d^3N_g^{(0)}}{dx \, d^2\textbf{k}_\perp} \right) \: ,
\end{align}
\end{subequations}
giving the vacuum contributions as
\begin{subequations}    \label{e:vacasym}
\begin{align}  
    \langle N_g \rangle_0 &\approx \frac{\alpha_s C_R}{\pi} \ln^2 \frac{E}{\mu}
    \\  \label{e:vacasymeloss}
    \left. \frac{\Delta E}{E} \right|_{\text{NLO, }0} &\approx \frac{2 \alpha_s C_R}{\pi} \ln\frac{E}{\mu} \: ,
    \\
    \langle p_T^2 \rangle_{\text{NLO, }0} &\approx 2 \frac{\alpha_s C_R}{\pi} E^2
\end{align}
\end{subequations}
in the high-energy limit at leading-logarithmic or leading-power accuracy.  Here we have integrated $0 \leq x \leq 1$ and $0 \leq k_\bot \leq 2 x E$ and set the scale of the dimensionless logarithm to be $\mu\sim T$ at this accuracy (later we will take $\mu$ to be the Debye mass of the medium).

%
\subsubsection{DGLV Energy Loss (\texorpdfstring{1$^{st}$}{1st} Order in Opacity)}
%

In this section, we present the Djordjevic-Gyulassy-Levai-Vitev (DGLV) formalism for the radiative energy loss.  Gyulassy et al.\ (GLV) \cite{Gyulassy:2000er} computed the all orders in opacity $(L/\lambda_g)^n$ expansion of the radiative energy loss for a fast massless parton in the QCD medium in the soft ($x\ll 1$) and collinear gluon emission ($\ltvec k\ll xE$) limits. Djordjevic and Gyulassy \cite{Djordjevic:2003zk} later generalized the massless result of GLV to derive the heavy quark medium-induced radiative energy loss to all orders in opacity $(L/\lambda_g)^n$, also for $x\ll 1$ with $\ltvec k\ll xE$. Djordjevic and Gyulassy's work involved the generalization of the GLV opacity series \cite{Gyulassy:2000er} to include massive quark kinematic effects and the inclusion of the Ter-Mikayelian plasmon effects for gluons discussed in \cite{Djordjevic:2003be}.  These calculations assumed the radiation to be soft and co\-llin\-e\-ar; that is, the fraction $x$ of energy carried away by an emitted gluon and the angle $k_\bot / x E$ at which it is emitted are both small. Additionally, it was assumed that the energy $E$ of the parton was the largest scale in the problem (the eikonal approximation).

%
\begin{figure*}
    \centering
  \includegraphics[width=\textwidth]{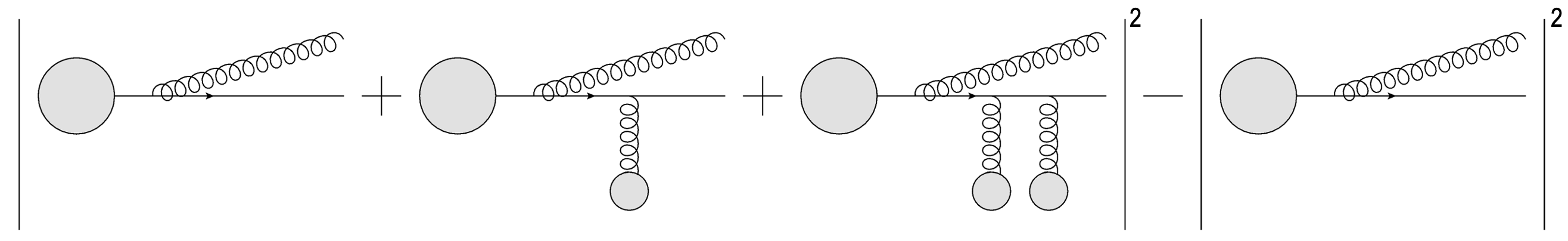}
  \caption{\label{fig:radiativefirst}
 Schematic depiction of the amplitude squared contributing to the first order in opacity contribution to the single inclusive radiative gluon emission spectrum; the diagrams corresponding to one or more interactions with the medium are representative of all possible attachments of the gluon exchange from the medium.  The vacuum contribution is explicitly removed.}
\end{figure*}
%

\fig{fig:radiativefirst} shows schematically the amplitude squared used to compute the medium induced single inclusive radiative gluon spectrum.  Since the vacuum emissions are explicitly subtracted out, the resulting single inclusive distribution can be positive or negative; when negative, the destructive interference from the LPM effect is dominant, and the amount of emitted radiation is \textit{less} than in vacuum. 

The resulting distribution of gluon radiation at first order in opacity generalizes the vacuum expression \eqref{eqn:dNvac} to be differential in the collisional momentum transfer $\tvec{q}$ with the medium, giving the kernel
\begin{align}\label{eqn:dNdxdkperp}
&\frac{d^5N_g^{(1)}}{dx d^2\textbf{k}_\perp d^2\textbf{q}_\perp} 
= \frac{1}{\pi} \frac{d^5N_g^{(1)}}{dx d\textbf{k}_\perp^2 d^2\textbf{q}_\perp} \nonumber\\
&= \frac{C_R \alpha_s}{\pi^3 x}\frac{L}{\lambda} \,  \frac{1}{\textbf{k}_\perp^2+m_g^2+M^2x^2} \frac{\mu^2}{(\textbf{q}_\perp^2+\mu^2)^2} \times \\
&\times 2 \frac{\textbf{k}_\perp\cdot \textbf{q}_\perp(\textbf{k}_\perp-\textbf{q}_\perp)^2 + (m_g^2+M^2x^2)\textbf{q}_\perp\cdot(\textbf{q}_\perp-\textbf{k}_\perp)}{(\frac{4Ex}{L})^2+ ((\textbf{k}_\perp-\textbf{q}_\perp)^2+M^2x^2+m_g^2)^2} \, .\nonumber
\end{align}
Note that the expression for \eqref{eqn:dNdxdkperp} assumes an exponentially falling distribution $ \sim \exp(-\Delta z/L)/L$ between the jet production and target rescattering center.  This simplified model of the medium was employed previously in Ref.~\cite{Gyulassy:2000er} to \linebreak smooth out the LPM interference pattern, roughly mimic the medium expansion, and permit an analytic expression for the distribution \eqref{eqn:dNdxdkperp} in closed form.  The details about the assumptions of the medium geometry are inessential for the qualitative comparison we wish to make, so it suffices for us to employ the same exponential model here.  For a quantitative comparison between formalisms it will be important to implement the same model of the medium in the twist-4 side as well.

As with the vacuum case \eqref{e:vackern}, the distribution \eqref{eqn:dNdxdkperp} of medium-induced radiation serves as the kernel for computing both the energy loss and momentum broadening.  Together, this gives
\begin{subequations}    \label{e:medkern}
\begin{align}   
    &\langle N_g \rangle_1 \equiv \int dx \, d^2 \tvec{k} \, d^2 \tvec{q} \:
    \frac{d^5 N_g^{(1)}}{dx \, d^2\textbf{k}_\perp \, d^2 \tvec{q}} \: .
    \\  \label{e:ElossNLO}
    &\left. \frac{\Delta E}{E} \right|_{\text{NLO, }1} =
    \int dx \, d^2 \tvec{k} \, d^2 \tvec{q} \: x \, \frac{d^5 N_g^{(1)}}{dx \, d^2\textbf{k}_\perp \, d^2 \tvec{q}}  \: ,
    \\  \label{eqn:pT2NLO}
    &\langle p_T^2 \rangle_{\text{NLO, }1} =
    \int dx \, d^2 \tvec{k} \, d^2 \tvec{q} \:  (\tvec{k} - \tvec{q})^2 \, \frac{d^5N_g^{(1)}}{dx \, d^2\textbf{k}_\perp \, d^2 \tvec{q}} \: ,
\end{align}
\end{subequations}

Writing the expression for the medium-induced, radiative energy loss out completely, we have
\begin{widetext}
    \begin{equation}\label{eqn:dEdximplement}
    \begin{split}
    \left. \frac{\Delta E}{E} \right|_{\text{NLO, }1}
    &= \frac{4C_R \alpha_s}{\pi^2}\frac{L}{\lambda} 
    \int_0^1 dx \,
    \int_0^{k_{\text{max}}} \frac{kdk}{k^2+m_g^2+M^2x^2}  \int_0^{q_{\text{max}}} qdq \,
    \frac{\mu^2}{(q^2+\mu^2)^2} \times 
    \\
    &\times  
    \int_0^{2\pi} d\theta \,
    \frac{kq\cos{\theta}(k^2+q^2-2kq\cos{\theta}) + (m_g^2+M^2x^2)(q^2-kq\cos{\theta})}{(\frac{4Ex}{L})^2+ ((k^2+q^2-2kq\cos{\theta})+M^2x^2+m_g^2)^2} \, ,
    \end{split}
    \end{equation}  
\end{widetext}
where we have defined $k\equiv|\textbf{k}_\perp|$ and $q\equiv|\textbf{q}_\perp|$. Note that $\theta$ measures the angle between $\textbf{k}_\perp$ and $\textbf{q}_\perp$. The kinematic limit
\begin{align}
    \label{e:lims}
    k_{\text{max}}&=2x(1-x)E \: 
\end{align}
is obtained by imposing collinearity on the emitted gluon, yielding $\ltvec k \lesssim 2xE$, and collinearity on the parent parton, yielding $\ltvec k \lesssim 2(1-x)E$.  Recalling that $x \ll 1$, it is sufficient to approximate Eq.~\eqref{e:lims} by $k_\mathrm{max} = 2 x E$ to determine the asymptotic scaling. 

\begin{figure}
    \centering
    \includegraphics[width=0.9\linewidth]{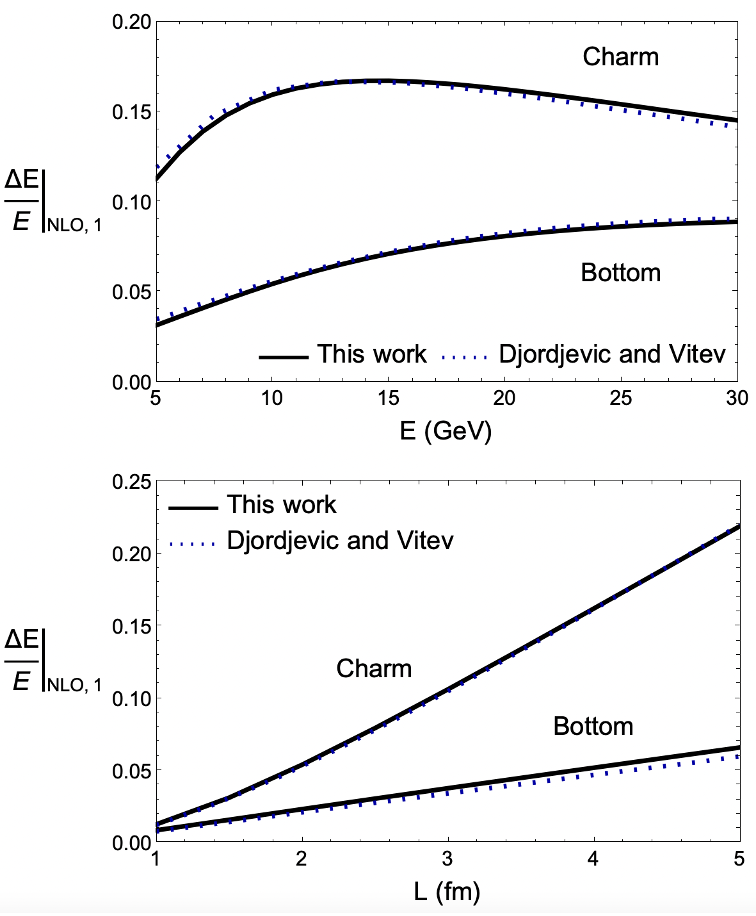}
    \caption{The first order in opacity fractional energy loss for charm and bottom quarks given by \eqref{eqn:dEdximplement} is plotted as a function of parent parton energy (upper plot) and the effective static thickness $L$ (lower plot) of a plasma characterised by $\alpha_s=0.3$, $\mu=0.5$ GeV and $\lambda=1$ fm. The black solid curves show the implementation of this work, while the blue dashed curves are given by the results of Djordjevic and Gyulassy (see Figures 1 and 2 in \cite{Djordjevic:2003zk}). For the upper plot we take $L=4$ fm, while the lower plot takes $E=10$ GeV. The charm and bottom quark masses are taken to be $m_c=1.6$ GeV and $m_b=4.75$ GeV respectively.}
    \label{fig:Eloss}
\end{figure}

To illustrate the behavior of the radiative energy loss given by \eqref{eqn:dEdximplement}, we plot the fractional energy loss for a charm and bottom quark as a function of initial parton energy $E$, and as a function of QGP effective length $L$ for fixed $E=10$ GeV, both in Figure~\ref{fig:Eloss}. The implementation in this work is compared to the same plots produced by Djordjevic and Gyulassy (Figures 1 and 2 in \cite{Djordjevic:2003zk}). We find that our Eq.~\eqref{eqn:dEdximplement} reproduces well the calculation from the original work of Djordjevic and Gyulassy.  As expected, the fractional energy loss decreases with the quark mass $M$ and increases like $L^2$ at small $L$, softening to a linear dependence $\propto L$ at large $L$.

%
%

%
\subsection{Total (LO+NLO) Transverse Momentum Broadening}
%

Having quantitatively verified in Fig.~\ref{fig:Eloss} the agreement of the DGLV energy loss expression \eqref{eqn:dEdximplement} as obtained by using the distribution \eqref{eqn:dNdxdkperp} as the energy loss kernel in Eq.~\eqref{e:ElossNLO}, we next consider the DGLV radiative momentum broadening \eqref{eqn:pT2NLO} obtained from the same kernel.  The observable of interest is the \textit{difference} in momentum broadening due to the medium, which is experimentally determined by a comparison of jet broadening in heavy-ion collisions versus proton-proton collisions:
\begin{align}
    \Delta \langle p_T^2 \rangle_{\text{tot}} 
    &\equiv \big[ \langle p_T^2 \rangle_{\text{tot, AA}} \big] - \big[ \langle p_T^2 \rangle_{\text{tot, pp}} \big] \nonumber \\
    &=  \big[ P_{\text{no rad}}\langle p_T^2 \rangle_{\text{el},AA} + P_{\text{rad},AA}\langle p_T^2 \rangle_{\text{inel},AA} \big] \nonumber\\
    & \quad - \big[ P_{\text{rad},pp}\langle p_T^2 \rangle_{\text{inel},pp} \big] \nonumber\\
\end{align}
\begin{align}\label{eqn:pT2tot}
    & = e^{- \langle N_g \rangle_0 - \langle N_g \rangle_1 } \langle p_T^2 \rangle_{\text{LO, }1} \nonumber\\
    & \quad + (1-e^{- \langle N_g \rangle_0 - \langle N_g \rangle_1 }) \left( \langle p_T^2 \rangle_{\text{NLO, 0}} + \langle p_T^2 \rangle_{\text{NLO, 1}}  \right)
    \nonumber \\ &
    \quad - (1-e^{-\langle N_g \rangle_{0}}) \langle p_T^2 \rangle_{\text{NLO, 0}} 
    \, ,
\end{align}
where the different collisional and radiative terms are weight\-ed by the average number of gluon emissions $\langle N_g \rangle$ at the indicated orders in opacity.  The collisional broadening term $\langle p_T^2 \rangle_{\text{LO, }1}$ is weighted by the probability \textit{not} to radiate a gluon $e^{- \langle N_g \rangle_0 - \langle N_g \rangle_1 }$ up to first order in opacity.  The radiative broadening for both the vacuum and the medium-induced radiation are similarly weighted by the probability $1 - e^{- \langle N_g \rangle}$ at the corresponding accuracies.  We will next proceed to evaluate Eq.~\eqref{eqn:pT2tot} and its various contributions both numerically and analytically in a leading-logarithmic analysis.

%
\section{Asymptotic scaling: Numerical}
\label{sec:numerics}
%

%
\begin{figure}
    \centering
    \includegraphics[width=\columnwidth]{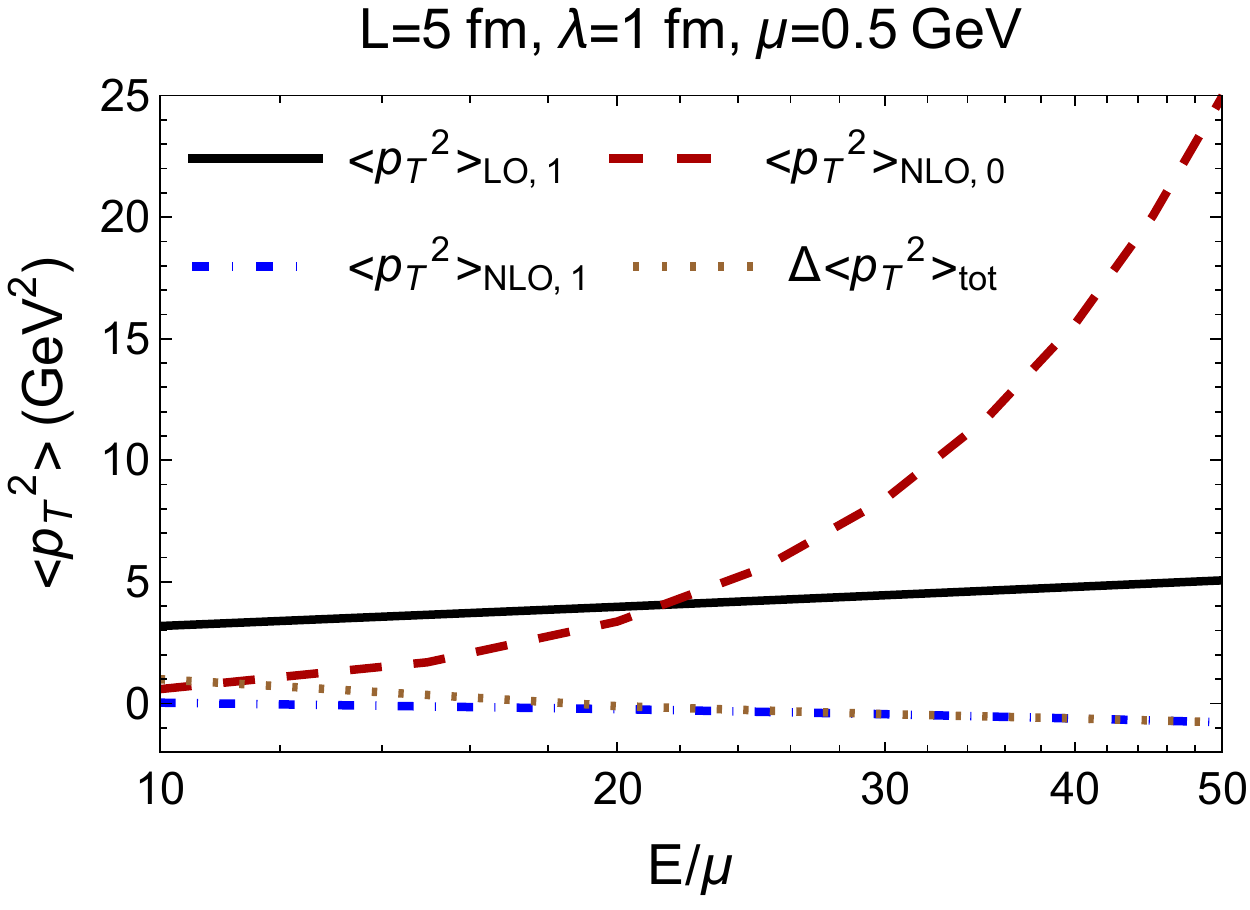}
    \caption{The LO, NLO, and vacuum-subtracted total transverse momentum picked up by a charm quark ($M=1.5$  GeV) propagating through a plasma characterized by $L=5\lambda=1$ fm and $\mu=0.5$ GeV. We observe $\ln(E/\mu)$ and $(\ln(E/\mu))^2$ scaling for the LO broadening $\langle p_T^2\rangle_{\text{LO, 1}}$ and first order in opacity NLO broadening $\langle p_T^2\rangle_{\text{NLO, 1}}$ respectively.}\label{fig:CharmBroadening}
\end{figure}
%

In this section, we demonstrate the numerical implementation of LO \eqref{eqn:pT2LO}, NLO \eqref{eqn:pT2LO} and total momentum broadening formulae \eqref{eqn:pT2tot} for massive and massless parent quarks from the opacity expansion energy loss formalism.  We focus in particular on the high-energy asymptotic limit $E \rightarrow \infty$ of this model, with the aim of comparing with the asymptotic behavior of the twist-4 calculation of Refs.~\cite{Kang:2014ela,Xing:2014kpa, Kang:2013raa}.  There, in the twist expansion, the radiative broadening effect (NLO) appears as a quantum evolution effect associated with large logarithms of $Q^2 / \mu_F^2$ with $\mu_F$ the factorization scale.  The result indicates that the radiative (NLO) and collisional (LO) broadening should be related by one step of logarithmic evolution, such that their ratio is proportional to the resummation parameter of the evolution equation: 
\begin{align}   \label{e:guess}
    \frac{\langle p_T^2 \rangle_\theRIGHTwayNLOone}{\langle p_T^2 \rangle_{\theRIGHTwayLOone}} \propto
    \left( \alpha_s \ln\frac{E}{\mu} \right) \: ,
\end{align}
with some constant of proportionality to be determined.

Unless otherwise specified, the broadenings defined by \eq{eqn:pT2LO}, \eq{eqn:pT2NLO} and \eq{eqn:pT2tot} are computed using the upper integration limits \eq{eq:qmax} and \eqref{e:lims}.  In Figure~\ref{fig:CharmBroadening}, we consider the LO, NLO (at zeroth and first order in opacity), and vacuum-subtracted total broadening of a heavy charm quark with mass $M=1.5$ GeV.  We observe that the vacuum broadening scales quadratically with the energy $\sim E^2$, consistent with Eq.~\eqref{e:vacasym}.  Similarly, the collisional (LO) broadening appears to scale like $\ln(E/\mu)$, consistent with Eq.~\eqref{e:CollAsym}.  Finally, the radiative (NLO) broadening at first order in opacity appears to scale like $\ln(E/\mu)^2$, consistent with the expectation \eqref{e:guess} of a single step of quantum evolution relative to the collisional broadening.  

\begin{figure}
    \centering
    \includegraphics[width=0.9\columnwidth]{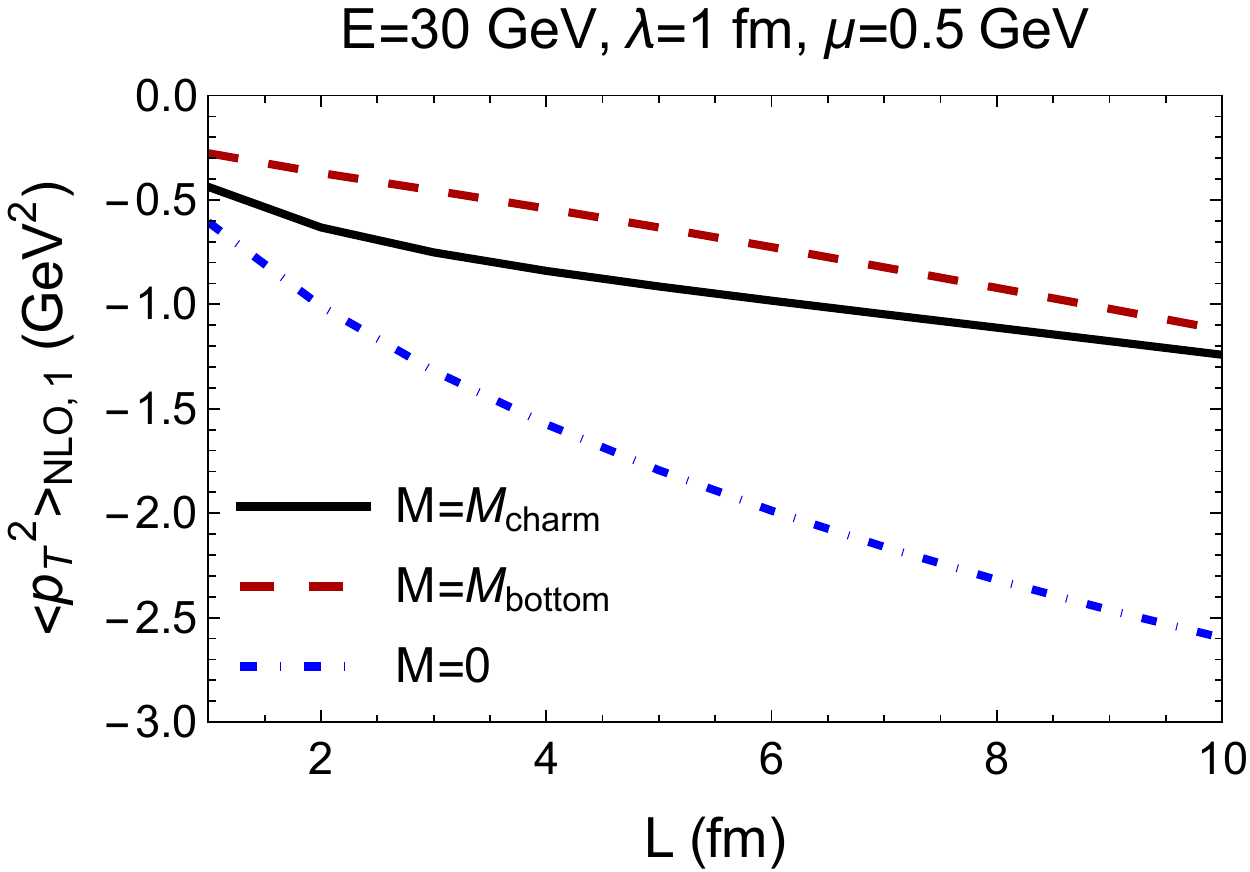}
    \caption{The radiative broadening of charm (solid), bottom (dashed), and massless (dot-dashed) quarks at first order in opacity, as a function of the effective medium length $L$. The charm and bottom quark masses were taken to be $M=1.5$ GeV and $M=4.75$ GeV respectively.}\label{fig:NLO_ML}
\end{figure}

\begin{figure}
    \centering
    \includegraphics[width=0.9\columnwidth]{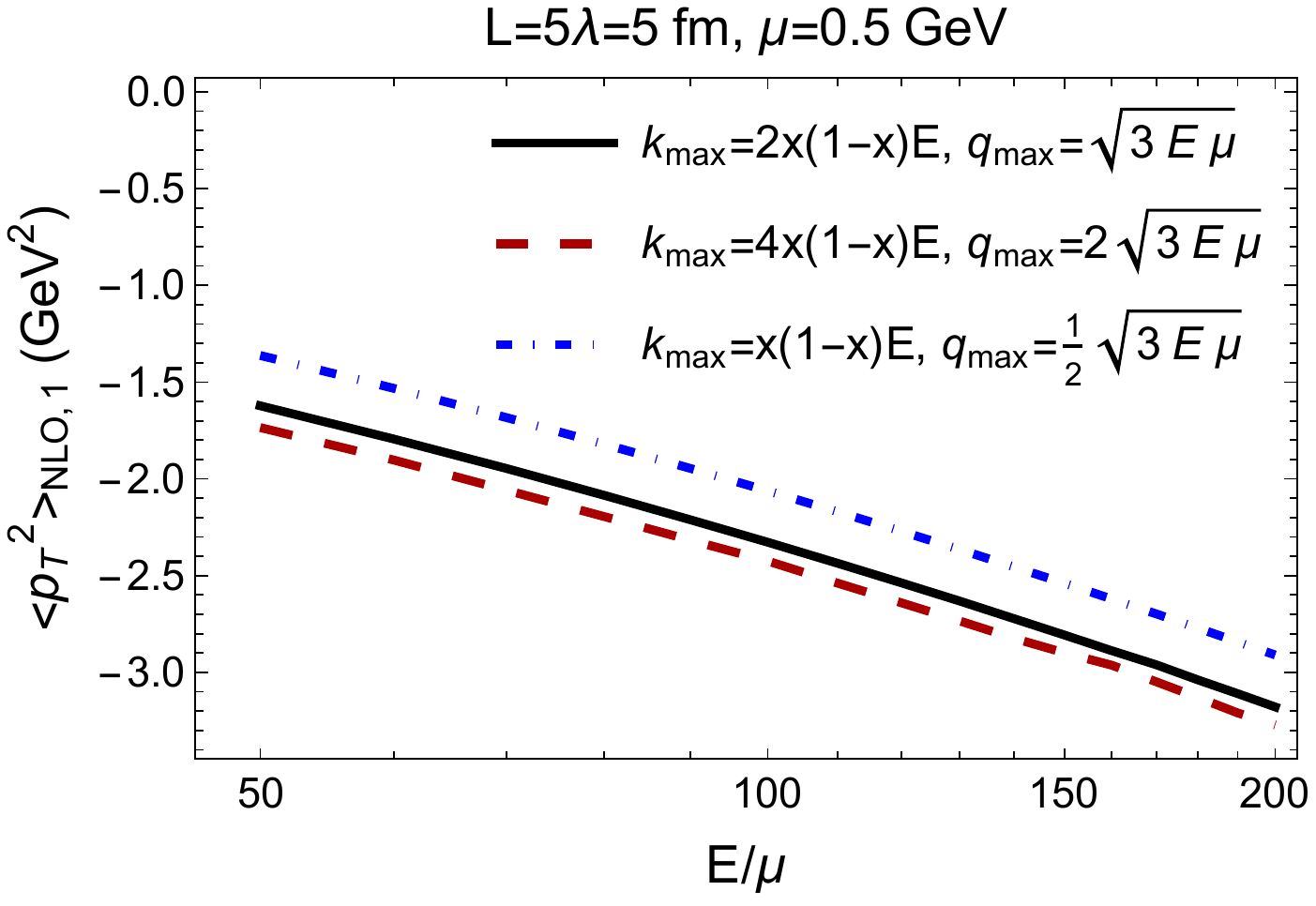}
    \caption{DGLV radiative broadening contribution at first order in opacity for different choices of the kinematic limits of integration.}
    \label{fig:limits}
\end{figure}

\begin{figure}
    \centering
    \includegraphics[width=0.9\columnwidth]{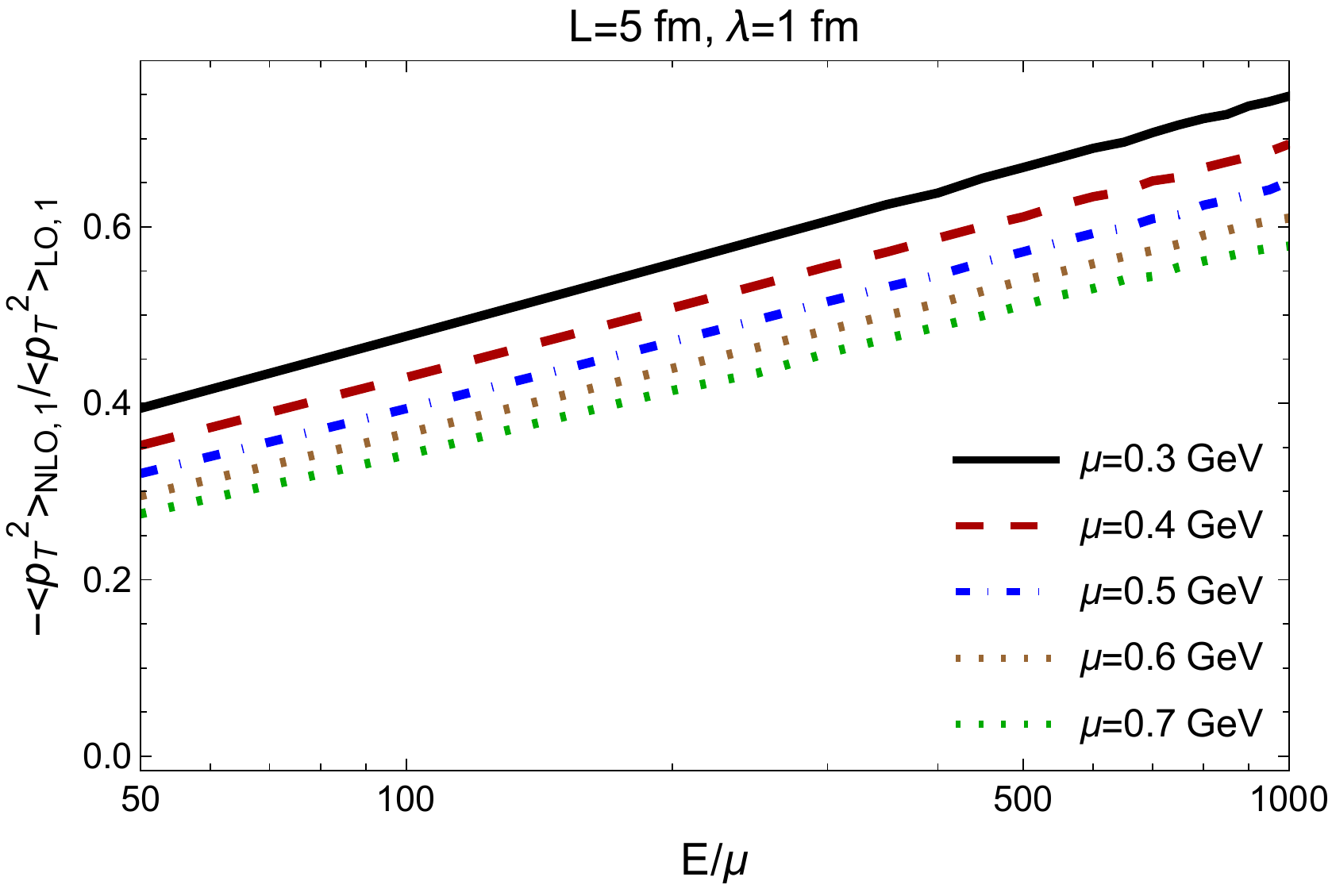}
    \caption{The ratio of (first-order) NLO to LO broadening of a massless parton as a function of $E/\mu$, for various values of $\mu$. For all values of $\mu$ considered, we observe a clear negative logarithmic dependence of the ratio on $E/\mu$.}\label{fig:ratioEoMu}
\end{figure}

However, a close examination of the radiative broadening component in Fig.~\ref{fig:CharmBroadening} reveals that the coefficient of proportionality anticipated in Eq.~\eqref{e:guess} is in fact \textit{negative}.  This negativity, implying a \textit{narrowing}, rather than a broadening, of the average transverse momentum when compared to the vacuum distribution, is robust for $\ord{1}$ variations in the parameters of the calculation, as shown dramatically in Figs.~\ref{fig:NLO_ML}, \ref{fig:limits}, and \ref{fig:ratioEoMu}.  Such a result, while counterintuitive, is a indeed a reasonable physical outcome of the LPM effect, which is in general a \textit{destructive} interference between the vacuum and medium-induced radiation.  While the LPM effect adds a net positive contribution to the energy loss of jets, the nontrivial redistribution of the radiated gluons leads to a net \textit{reduction} of the radiative broadening compared to vacuum.

%
%

For heavy quarks (charm and bottom), we see explicitly the simultaneous increase in radiative energy loss (Fig.~\ref{fig:Eloss}) and decrease in radiative broadening (Fig.~\ref{fig:NLO_ML}).  Both effects show a mass ordering, with the heavier bottom quarks losing less energy and narrowing less compared to the charm quarks.  Massless quarks lose the most energy \cite{Gyulassy:2000er} and are narrowed the most.  Variations in the precise choice of the kinematic limits shown in Fig.~\ref{fig:limits} do not change the qualitative narrowing of the momentum distribution, shown here for the massless case.  Moreover, the rate of this logarithmic growth with $E$ (slope of the curves in Fig.~\ref{fig:limits}) is independent of the precise values of the limits as well.  

%
%

%
%

Focusing on the massless limit ($M=m_g=0$) to better study the high-energy asymptotics, we plot the absolute value of the ratio $\langle p_T^2 \rangle_{\text{NLO, 1}} / \langle p_T^2 \rangle_{\text{LO, 1}}$ in Fig.~\ref{fig:ratioEoMu} to study the coefficient of the logarithm anticipated by Eq.~\eqref{e:guess}.  In this limit the distribution \eqref{eqn:dNdxdkperp} simplifies to the original expression of Gyulassy et al.\ for massless partons \cite{Gyulassy:2000er} 
\begin{equation}\label{eqn:dNdxdkperpGLV}
\begin{split}
\frac{d^5N_g^{(1)}}{dx d^2\textbf{k}_\perp d^2\textbf{q}_\perp} 
&= \frac{C_R \alpha_s}{\pi^3 x}\frac{L}{\lambda} \, \frac{1}{\textbf{k}_\perp^2} \frac{\mu^2}{(\textbf{q}_\perp^2+\mu^2)^2} \\
&\times 2 \frac{\textbf{k}_\perp\cdot \textbf{q}_\perp(\textbf{k}_\perp-\textbf{q}_\perp)^2 L^2}{(4Ex)^2+ (\textbf{k}_\perp-\textbf{q}_\perp)^4 L^2} \, .
\end{split}
\end{equation}
The linearity of the curves shown in Fig.~\ref{fig:ratioEoMu} confirms that the scaling is indeed logarithmic as expected.  However, the negativity of the constant of proportionality is concerning.  If it is indeed the true prediction of energy loss approaches such as DGLV, then we would like to substantiate that by an explicit analytic evaluation of the high-energy asymptotics.  We perform this analysis next in Sec.~\ref{sec:analytics} and compare with the asymptotic high-energy behavior seen in the direct numerical evaluation.

%
\section{Asymptotic scaling: Analytic}
\label{sec:analytics}
%

As we shown strikingly in Figs.~\ref{fig:CharmBroadening} and \ref{fig:ratioEoMu}, a direct numerical evaluation of the radiative (NLO) broadening in the DGLV formalism predicts a \textit{narrowing} of the transverse momentum distribution in medium.  To understand the origin of this asymptotic behavior, we undertake in this section an analytic treatment of the high-energy limit in various ways.  First we will consider a simplified (but standard) treatment of the high-energy kinematics, which ignores finite kinematic bounds and integrates the distributions to infinity.  This procedure is known to provide an accurate asymptotic estimate of the \textit{energy loss}, but as we will show, fails qualitatively for the transverse momentum broadening.  The origin of this discrepancy is also relevant to comparing between energy loss frameworks such as DGLV and the collinear twist-4 approach, so it is instructive to begin there as a baseline for comparison.  Then we will re-examine the high-energy asymptotics by carefully treating the finite kinematic bounds.  To ensure the accuracy of the analytics, we will benchmark all necessary approximations with direct numerical comparison.  The final result of our analysis does qualitatively change the sign of the effect, predicting a net narrowing as seen in the exact numerics.

%
\subsection{Infinite Kinematics Approximation}
\label{sec:infiniteasymptotics}
%

%
\subsubsection{As Applied to Radiative Energy Loss}
\label{sec:infeloss}
%

The calculation of the fractional parton energy loss $\Delta E / E$ in the vacuum ($\mathrm{NLO}, \: 0$) and at first order in opacity ($\mathrm{NLO}, \: 1$) was performed previously in Ref.~\cite{Gyulassy:2000er}.  Here we summarize their method, which simplifies the kinematic bounds to make the results more analytically tractable, before applying the same logic to the calculation of the momentum broadening.  The vacuum result \eqref{e:vacasymeloss} is trivially obtained from Eqs.~\eqref{e:vackern} by integrating $k_\bot \in [0, 2 x E]$.  Note that the result \eqref{e:vacasymeloss} we show here differs from Eq.~(122) of Ref.~\cite{Gyulassy:2000er} by a factor of $\tfrac{2}{3}$ due to their retention of the full polynomial $1 - x + \tfrac{x^2}{2}$, which we approximate as $1$ under the condition $x \ll 1$ under which it was derived.

To perform a similar calculation of the energy loss $\Delta E / E$ at first order in opacity from \eq{e:ElossNLO}, we follow the calculation performed in Ref.~\cite{Gyulassy:2000er}.  First we change integration variables to define 
\begin{align}
    \qprime \equiv \tvec{k} - \tvec{q}
\end{align}
in terms of which the differential distribution \eqref{eqn:dNdxdkperpGLV} becomes
\begin{align}   \label{e:distribn2}
    &\frac{dN^{(1)}}{dx \, d^2 \tvec{k} \, d^2 \tvec{q}^\prime} =
    \notag \\ &=
    \frac{C}{x} \: \frac{\tvec{k} \cdot (\tvec{k} - \qprime)}{k_\bot^2}
    \frac{\mu^2}{((\tvec{k} - \qprime)^2 + \mu^2)^2}
    \frac{q_\bot^{\prime \: 2}}{A^2 + q_\bot^{\prime \: 4}}
    \notag \\ &=
    \frac{C}{x} \: \frac{\mu^2}{k_\bot^2}
    \frac{
        k_\bot^2 - k_\bot q_\bot^\prime \cos\theta
        }{
        (k_\bot^2 + q_\bot^2 - 2 k_\bot q_\bot^\prime \cos\theta + \mu^2)^2
        }
    \frac{q_\bot^{\prime \: 2}}{A^2 + q_\bot^{\prime \: 4}}
\end{align}
with unit Jacobian.  Here $\theta$ is the angle between $\tvec{k}$ and $\qprime$, and for brevity we have introduced the coefficients 
\begin{subequations}
\begin{align}
    A &= \frac{4 x E}{L}    \\
    C &= 2 \frac{\alpha_s C_R}{\pi^3} \frac{L}{\lambda} = \frac{8 \alpha_s}{3 \pi^3} \frac{L}{\lambda} \: .
\end{align}
\end{subequations}

To study the high-energy asymptotics, Gyulassy et al. integrated the kernel \eqref{e:distribn2} as in \eq{e:ElossNLO} over the entire range from $q_\bot^\prime , k_\bot \in [0,\infty)$, giving
\begin{align}   \label{e:eloss1}
    &\left. \frac{\Delta E}{E} \right|_{\text{NLO, }1} =
    \frac{C \pi}{2} \int\limits_{0}^{1} dx
    \int\limits_{0}^{\infty} dk_\bot^2
    \int\limits_{0}^{2\pi} d\theta
    \int\limits_{0}^{\infty} dq_\bot^{\prime \: 2} \:
    \notag \\ &\times
    \frac{\mu^2}{k_\bot^2}
    \frac{
        k_\bot^2 - k_\bot q_\bot^\prime \cos\theta
        }{
        (k_\bot^2 + q_\bot^2 - 2 k_\bot q_\bot^\prime \cos\theta + \mu^2)^2
        }
    \frac{q_\bot^{\prime \: 2}}{A^2 + q_\bot^{\prime \: 4}} \: .
\end{align}
We observe that the numerator $k_\bot^2 - k_\bot q_\bot^\prime \cos\theta$ from \eq{e:eloss1} can be simply obtained by differentiation of the denominator:
\begin{align}
    &\frac{k_\bot^2 - k_\bot q_\bot^\prime \cos\theta}
    {(k_\bot^2 + q_\bot^2 - 2 k_\bot q_\bot^\prime \cos\theta + \mu^2)^2} =
    \notag \\ &=
    \label{eq:deriv}
    - \frac{k_\bot}{2} \frac{\partial}{\partial k_\bot}
    \left(
    \frac{1}{k_\bot^2 + q_\bot^2 - 2 k_\bot q_\bot^\prime \cos\theta + \mu^2}
    \right) \: .
\end{align}
\eq{eq:deriv} allows us to perform the $d\theta$ integration first over the simpler integrand, then differentiate back afterwards:
\begin{align}   \label{e:eloss2}
    \left. \frac{\Delta E}{E} \right|_{\text{NLO, }1} &=
    C \pi^2 \mu^2 \int\limits_{0}^{1} dx
    \int\limits_{0}^{\infty} dq_\bot^{\prime \: 2} \:
    \int\limits_{\mu^2}^{\infty} dk_\bot^2 \:
    \frac{q_\bot^{\prime \: 2}}{A^2 + q_\bot^{\prime \: 4}}
    \notag \\ &\times
    \frac{ k_\bot^2 - q_\bot^{\prime \: 2} + \mu^2}{
    \left[ k_\bot^4 - 2 k_\bot^2 (q_\bot^{\prime 2} - \mu^2) + (q_\bot^{\prime 2} + \mu^2)^2 \right]^{3/2}
    } \: .
\end{align}
Then the $d k_\bot^2$ integral can be performed analytically, giving
\begin{align}   \label{e:eloss3}
    \left. \frac{\Delta E}{E} \right|_{\text{NLO, }1} &=
    C \pi^2 \mu^2 \int\limits_{0}^{1} dx
    \int\limits_{0}^{\infty} dq_\bot^{\prime \: 2} \:
    \frac{q_\bot^{\prime \: 2}}{A^2 + q_\bot^{\prime \: 4}} \:
    \frac{1}{q_\bot^{\prime \: 2} + \mu^2} \: ,
\end{align}
and similarly for the $dq_\bot^{\prime \: 2}$ integral:
\begin{align}   \label{e:eloss4}
    \left. \frac{\Delta E}{E} \right|_{\text{NLO, }1} &=
    C \pi^2 \mu^2 \int\limits_{0}^{1} dx
    \frac{A \pi - 2 \mu^2 \ln\frac{A}{\mu^2}}{2(A^2 + \mu^4)}
    \notag \\ &=
    C \pi^2 \int\limits_{0}^{1} dx
    \frac{
        \left( \frac{4 x E}{\mu^2 L} \right) \pi 
        - 2 \ln\left(\frac{4 x E}{\mu^2 \, L}\right)
    }{
        2\left[ \left( \frac{4 x E}{\mu^2 L} \right)^2 + 1 \right]
    }
    \: .
\end{align}
Assuming that the dominant limit is
\begin{align}
    \frac{4 x E}{\mu^2 L} \gg 1
    \qquad \rightarrow \qquad
    x \gg \frac{\mu^2 L}{4 E} \,
\end{align}
the remaining $dx$ integral becomes logarithic
\begin{align}   \label{e:eloss5}
    \left. \frac{\Delta E}{E} \right|_{\text{NLO, }1} &\approx
    C \pi^3 \: \frac{\mu^2 L}{8 E}
    \int\limits_{0}^{1} \frac{dx}{x} \: .
\end{align}
This last integral must be regulated in the small-$x$ regime with $x \gg \ord{\frac{\mu}{E}}$, giving
\begin{align}   \label{e:eloss6}
    \left. \frac{\Delta E}{E} \right|_{\text{NLO, }1} &\approx
    C \pi^3 \: \frac{\mu^2 L}{8 E} \: \ln\frac{E}{\mu}
    \notag \\ &=
    \frac{\alpha_s C_R}{4} \: 
    \frac{\mu^2 L^2}{\lambda \, E} \: \ln\frac{E}{\mu} \: .
\end{align}
Comparing the high-energy asymptotics \eqref{e:eloss6} to the vacuum energy loss \eqref{e:vacasymeloss}, we see that the medium-induced contribution is suppressed compared to the vacuum by a factor
\begin{align}   \label{e:eloss7}
    \left. \frac{\Delta E}{E} \right|_{\text{NLO, }1} =
    \left( \frac{\pi}{8} \: 
    \frac{\mu^2 L^2}{\lambda \, E} \right)
    \: 
    \left. \frac{\Delta E}{E} \right|_{\text{NLO, }0} \: ,
\end{align}
indicating that the LPM effect has softened the medium-induced energy loss relative to the vacuum.

%
\subsubsection{As Applied to Radiative Broadening}
\label{sec:infbroad}
%

Next we want to apply the same logic to the calculation of the radiative momentum broadening \eqref{eqn:pT2LO} in the medium.  As before, we will change variables to $\qprime \equiv \tvec{k} - \tvec{q}$, but we must take additional care with the limits because of the different weighting of the $x$ and $q_\bot^\prime$ integrals.  To that end, let us determine the appropriate kinematic bounds on the $\qprime$ integration corresponding to Eqs.~\eqref{e:lims}.  The upper limit of the $q_\bot^\prime < q_{max}^\prime$ integration is determined by the condition
\begin{align}
    (q_\bot^2)_{max} = k_\bot^2 + q_{max}^{\prime \: 2} - 2 k_\bot q_{max}^\prime \cos\theta = 3 E \mu \: .
\end{align}
This quadratic equation in $q_{max}^\prime$ has the two solutions
\begin{align}
    q_{max}^\prime = k_\bot \cos\theta \pm \sqrt{3 E \mu - k_\bot^2 \sin^2 \theta}
\end{align}
which define the inner / outer boundaries of the $\qprime$ integration region.  The (log-divergent) large-$E$ behavior is governed by the large phase space of the $\qprime$ integration; as we will show, the leading behavior of $q_{max}^{\prime \: 2}$ with $E$ is what generates the leading \textit{double}-logarithmic behavior.  Subleading corrections which scale like $\sqrt{E}$ may contribute to single-logarithmic corrections which are higher-order, and terms which are finite in $E$ can never generate divergences as $E \rightarrow \infty$.  With this analysis, we conclude that the leading behavior of the $\qprime$ phase space as $E \rightarrow \infty$ is
\begin{align}
    q_{max}^{\prime \: 2} \approx 3 E \mu 
\end{align}
which is nicely independent of the other variables $k_\bot$ and $\theta$ which enter into the integral \eqref{eqn:pT2NLO}.  It is only because the leading upper limit is a constant independent of $k_\bot$ and $\theta$ that the machinery developed in Ref.~\cite{Gyulassy:2000er} can be applied in the same form.  The radiative momentum broadening \eqref{eqn:pT2NLO} is then given by
\begin{align}   \label{e:broad2}
    &\langle p_\bot^2 \rangle_{\text{NLO, }1} =
    \frac{C \pi}{2} \int\limits_{\frac{\mu}{E}}^{1} \frac{dx}{x}
    \int\limits_{\mu^2}^{4 x^2 E^2} dk_\bot^2
    \int\limits_{0}^{2\pi} d\theta
    \int\limits_{0}^{3 E \mu} dq_\bot^{\prime \: 2} \:
    \notag \\ &\times
    \frac{\mu^2}{k_\bot^2}
    \frac{
        k_\bot^2 - k_\bot q_\bot^\prime \cos\theta
        }{
        (k_\bot^2 + q_\bot^2 - 2 k_\bot q_\bot^\prime \cos\theta + \mu^2)^2
        }
    \frac{q_\bot^{\prime \: 4}}{A^2 + q_\bot^{\prime \: 4}} \: ,
\end{align}
where integrating over the angle of the absolute coordinate axes gave a factor of $(2\pi)$ and changing variables from $k_\bot dk_\bot$ to $dk_\bot^2$ (and similarly for $q'$) generated a factor of $(1/2)^2$.

Because the leading upper limit of $q_\bot^\prime$ is independent of the angle $\theta$, we can perform the angular integration of \eq{e:broad2} using the derivative technique from before.  Writing the numerator $k_\bot^2 - k_\bot q_\bot^\prime \cos\theta$ from \eq{e:broad2} in terms of a derivative allows us to perform the angular integral, giving
\begin{align}   \label{e:broad3}
    \langle p_\bot^2 \rangle_{\text{NLO, }1} &=
    C \pi^2 \mu^2 \int\limits_{\frac{\mu}{E}}^{1} \frac{dx}{x}
    \int\limits_{0}^{3 E \mu} dq_\bot^{\prime \: 2} \:
    \int\limits_{\mu^2}^{4 x^2 E^2} dk_\bot^2 \:
    \frac{q_\bot^{\prime \: 4}}{A^2 + q_\bot^{\prime \: 4}}
    \notag \\ &\times
    \frac{ k_\bot^2 - q_\bot^{\prime \: 2} + \mu^2}{
    \left[ k_\bot^4 - 2 k_\bot^2 (q_\bot^{\prime 2} - \mu^2) + (q_\bot^{\prime 2} + \mu^2)^2 \right]^{3/2}
    } \: .
\end{align}
In the energy loss calculation, it was sufficient to extend the UV limit of the $dk_\bot^2$ integral to infinity, neglecting the finite kinematic bound $k_\bot < 2 x E$.  As we shall see, while this approximation sufficed for the case of energy loss, it becomes much more tenuous for the case of radiative momentum broadening.  Therefore, as a baseline for comparison, let us consider the result we obtain by similarly integrating $k_\bot^2$ all the way from $0$ to $\infty$ instead of over its appropriate finite bounds.  Then the $d k_\bot^2$ can be performed analytically, giving
\begin{align}   \label{e:broad4b}
    \langle p_\bot^2 \rangle_{\text{NLO, }1} &=
    C \pi^2 \mu^2 \int\limits_{\frac{\mu}{E}}^{1} \frac{dx}{x}
    \int\limits_{0}^{3 E \mu} dq_\bot^{\prime \: 2} \:
    \frac{q_\bot^{\prime \: 4}}{A^2 + q_\bot^{\prime \: 4}} \:
    \frac{1}{q_\bot^{\prime \: 2} + \mu^2} \: .
\end{align}
We note that the $dq_\bot^{\prime \:2}$ integral becomes logarithmic $\sim \frac{ dq_\bot^{\prime \:2} }{ q_\bot^{\prime \:2} }$ if we satisfy the two criteria
\begin{subequations}    \label{e:crits1}
\begin{align}
    q_\bot^{\prime \: 2} &\gg A = \frac{4 x E}{L} \\
    q_\bot^{\prime \: 2} &\gg \mu^2 \: .
\end{align}
\end{subequations}
When both of these criteria are satisfied $q_\bot^{\prime \: 2} \gg (\max[ A \, , \, \mu^2])$, the integral becomes logarithmic, but which of the two criteria \eqref{e:crits1} is the more restrictive depends on $x$.  For $x < x_c$ with the critical switching value of $x$ being
\begin{align}
    x_c = \frac{\mu^2 L}{4 E}
\end{align}
the maximum is $\mu^2$; if $x > x_c$ the maximum is $A = 4 x E / L$.  Splitting up the $x$ integral into these two regions gives
\begin{align}   \label{e:broad5}
    &\langle p_\bot^2 \rangle_{\text{NLO, }1} 
    \notag \\ &=
    C \pi^2 \mu^2 
    \left[
        \int\limits_{\frac{\mu}{E}}^{x_c} \frac{dx}{x}
        \ln\frac{3 E}{\mu}       
        + \int\limits_{x_c}^{1} \frac{dx}{x}
        \ln\frac{3 E \mu}{4 x E / L}
    \right]
    \notag \\ &\approx
    \frac{4 \alpha_s}{3 \pi} \, \mu^2\frac{L}{\lambda} \, \ln^2 \frac{E}{\mu} + \ord{\alpha_s \ln\frac{E}{\mu}}
    \: ,
\end{align}
where only the second integral over the parametrically large region $\frac{\mu^2 L}{4 E} < x < 1$ produces a double logarithm of the jet energy.  

Compared to the collisional broadening \eqref{e:CollAsym}, we see that the radiative broadening \eqref{e:broad5} is suppressed by a factor of $\alpha_s$, but enhanced by a logarithm of the jet energy:
\begin{align}   \label{e:DGLVratio}
    \frac{\langle p_\bot^2 \rangle_{\text{NLO, }1}}
    {\langle p_\bot^2 \rangle_{\text{LO, }1}}
     =
    \frac{4 \alpha_s}{3 \pi} \, \ln\frac{E}{\mu}
     \: .
\end{align}
This result is indeed compatible with the expectation from \eqref{e:guess} that the radiative broadening occurs as a quantum evolution correction to the collisional broadening.  Notably, the coefficient of the logarithmic, evolution-like correction is \textit{positive}, reflecting an increase in the broadening compared to the collisional term, in direct contrast to the \textit{narrowing} observed numerically in the previous section.

Finally we note that direct numerical evaluation of \linebreak \eq{e:broad2} is extremely delicate and must be handled with care.  The rapidly decaying and oscillating integrand is highly susceptible to numerical cancellation and roundoff error, so careful convergence and consistency tests of the numerics are essential.  Wherever possible, evaluating part of the expression analytically (as in the $dk_\bot^2$ integral performed in obtaining \eq{e:broad4b}) helps stabilize the numerics.

%
\subsection{Subtleties of Finite Kinematics}
\label{sec:finitebroad}
%

\subsubsection{Preliminaries}

In particular, we wish to evaluate \eq{eqn:pT2NLO} analytically for $\langle \Delta p_T^2\rangle$ from radiative energy loss in the limit of $E\gg\mu$.  We have not found an easy way to isolate the leading double logarithmic behavior of the integral.  This difficulty is likely related to the fact that, despite the superficially log IR divergent $1/x$ and linearly UV divergent $(\ltvec{k})^0$ nature of the integrand, the integral is actually convergent.  This convergence appears to be due to an extremely delicate cancellation of competing divergences.  (It's perhaps worth noting that even subleading corrections to the radiative energy loss kernel can destroy this delicate cancellation of divergences, changing the leading in $E$ behavior from $\Delta E\sim\ln E$ to $\Delta E\sim E$ \cite{Kolbe:2015rvk}.)  Possibly another way of seeing the difficulty of extracting the leading behavior is that we expect the leading contribution to come from $\tvec{k}\sim\tvec{q}$ with $4xE/L$ acting as a regulator; however we must also integrate over $x$, and, worse, $k_{max}\sim x$ for small $x$.  We were successful in extracting the leading in energy behavior only through brute force evaluation of the integral, then expanding for large energies.

The advantage of the form of the equation in \eq{eqn:pT2NLO} is that the domains of integration are especially simple: $x\in(0,1)$, $\ltvec{k}\in\big(0,k_{max}(x,E)\big)$, and $\ltvec{q}\in\big(0,q_{max}(\mu,E)\big)$.  In particular, there's no non-trivial angular dependence in any of the regions of integration.  The penalty is the difficult dependence on $(\tvec{q}-\tvec{k})^4$.  One way to make progress, as shown in the previous section, is to perform a change of variables to $\tvec{q'}\equiv\tvec{q}-\tvec{k}$.  The trade-off is that the domain of integration becomes significantly more complicated.  The $\tvec{q'}$ integration is still a disc of radius $q_{max}=\sqrt{3\mu E}$; however, this disc is now shifted away from the origin by a distance $\ltvec{k}$.  There are two possibilities: the $\tvec{q'}$ integration region continues to contain the origin ($\tvec{q'}=\tvec{0}$), what we'll from now on refer to as a ``small shift''; or the $\tvec{q'}$ integration region no longer contains the origin, what we'll from now on refer to as a ``large shift.''  Clearly the large shift only occurs when $\ltvec{k}$ is large; in particular, a large shift can only occur when $\ltvec{k}>q_{max}$.  We thus have that, after the shift in integration variables,
\begin{widetext}
    \begin{align}
        \langle\ltvec{p}^2\rangle_{\text{NLO, }1}&=
        \frac{C_R\alpha_s}{\pi^2} \frac L\lambda \int dx d^2\tvec{k}d^2\tvec{q}\frac1x \frac{\tvec{k}\cdot\tvec{q}}{\ltvec{k}^2}\frac{\mu^2}{(\ltvec{q}^2+\mu^2)^2}\frac{(\tvec{k}-\tvec{q})^4}{(4xE/L)^2+(\tvec{k}-\tvec{q})^4} \nonumber\\
        \label{eq:shiftedintegrand}
        & = \frac{2C_R\alpha_s}{\pi}\frac L\lambda \bigg\{ 
        \int_0^{x_{min}} dx\int_0^{k_{max}}d\ltvec{k}\int_0^{2\pi}d\theta\int_0^{q'_{max,+}}d\ltvec{q'} I(x,\ltvec{k},\ltvec{q},\theta,E,L,\mu) \nonumber\\
        & \qquad\qquad\quad + \int_{x_{min}}^1 dx\int_{0}^{q_{max}}d\ltvec{k}\int_0^{2\pi}d\theta\int_0^{q'_{max,+}}d\ltvec{q'} I(x,\ltvec{k},\ltvec{q},\theta,E,L,\mu) \\
        & \qquad\qquad\quad + \int_{x_{min}}^1 dx\int_{q_{max}}^{k_{max}}d\ltvec{k}\int_{-\theta_{max}}^{\theta_{max}}d\theta\int_{q'_{max,-}}^{q'_{max,+}}d\ltvec{q'} I(x,\ltvec{k},\ltvec{q},\theta,E,L,\mu)\bigg\}; \nonumber\\
        I & \equiv \frac{\ltvec{q'}(\ltvec{k}-\ltvec{q'}\cos\theta)}{x}\frac{\mu^2}{(\ltvec{k}^2+\ltvec{q}^{\prime2}-2\ltvec{k}\ltvec{q'}\cos\theta+\mu^2)^2}\frac{\ltvec{q}^{\prime4}}{(4xE/L)^2+\ltvec{q}^{\prime4}}
    \end{align}
\end{widetext}
where
\begin{align}
    x_{min} & \equiv \sqrt{\frac{3\mu}{4E}} \nonumber\\
    k_{max} & \equiv 2xE \nonumber\\
    q_{max} & \equiv \sqrt{3\mu E}\\
    \theta_{max} & \equiv \sin^{-1}\Big( \frac{q_{max}}{\ltvec{k}}\Big) \nonumber\\
    q'_{max,\pm} & \equiv \ltvec{k}\cos\theta\pm\sqrt{q^2_{max}-\ltvec{k}^2\sin^2\theta}.\nonumber
\end{align}
Note that $x_{min}$ is the solution to $k_{max}(x_{min},E)=q_{max}(\mu,E)$ such that for $x>x_{min}$ one has that $k_{max}>q_{max}$.  In \eq{eq:shiftedintegrand}, the first two lines correspond to the ``small shift'' while the third line corresponds to the ``large shift.''  Here we've taken $k_{max}=2xE$ instead of the usual $2x(1-x)E$ for simplicity.  The GLV formula is derived in the limit $x\ll1$, so using the simpler $k_{max}$ is consistent with the usual GLV approximations.

\begin{figure}
    \centering
    \includegraphics[width=0.9\columnwidth]{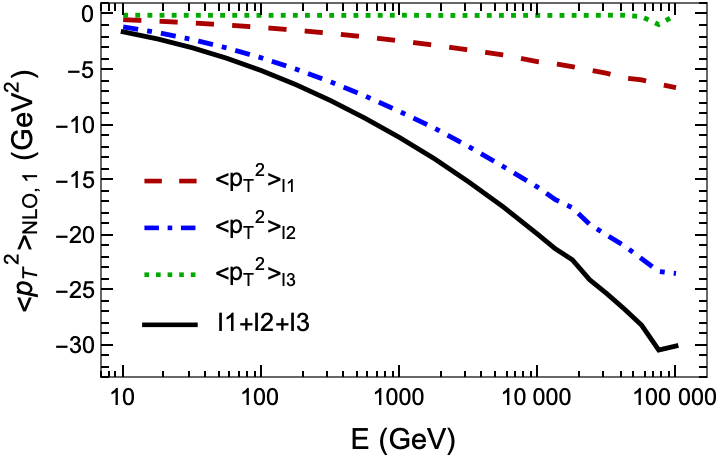}
    \caption{Comparison of the three contributions to $\langle\ltvec{p}^2\rangle_{\text{NLO, }1}$ from the three lines of \protect\eq{eq:shiftedintegrand} for $L=5$ fm and $\mu=0.5$ GeV: the small shift, small $x$ contribution from the first line of \protect\eq{eq:shiftedintegrand} (dashed); the small shift, large $x$ contribution from the second line of \protect\eq{eq:shiftedintegrand} (dash-dotted); the large shift contribution from the third line of \protect\eq{eq:shiftedintegrand} (dotted); and the sum of the three contributions (solid).}
    \label{fig:contributioncomparison}
\end{figure}

We show in \fig{fig:contributioncomparison} the three separate contributions from the three lines in \eq{eq:shiftedintegrand} to $\langle\ltvec{p}^2\rangle_{\text{NLO, }1}$.  One can see the onset of significant numerical instability for $E\gtrsim2000$ GeV.  As can be seen in the figure, in the high-energy limit, one may safely neglect the large shift integral as, at most, contributing a very small, approximately energy-independent amount.  We will therefore neglect the large shift integral from now on.  

Of the small shift integrals, there are two relevant regions: when $\ltvec{k}$ is cut off by $k_{max}$, the first line of \eq{eq:shiftedintegrand}, and when $\ltvec{k}$ is cut off by $q_{max}$, the second line of \eq{eq:shiftedintegrand}.  $\ltvec{k}$ is cut off by $k_{max}$ when $x<x_{min}$ and by $q_{max}$ otherwise; thus we refer to these two contributions as the small shift, small $x$ integral and the small shift, large $x$ integral.  \fig{fig:contributioncomparison} suggests that the two small shift contributions grow like $\ln^2(E)$, with the small shift, large $x$ contribution  about 3 times larger than the small shift, small $x$ contribution.   We will evaluate the small shift, large $x$ contribution first for two reasons: the small shift, large $x$ contribution is the larger of the two; and because $q_{max}$ doesn't depend on $x$, the small shift, large $x$ contribution is easier of the two to evaluate.

\subsubsection{Small Shift, Large \texorpdfstring{$x$}{x} Integral}
We'd like to examine the leading in energy behavior of the second line of \eq{eq:shiftedintegrand}.  
Let's first make the integral dimensionless by scaling out $\sqrt{3\mu E}$ from both $\ltvec{k}$ and $\ltvec{q'}$.  Defining $k\equiv \ltvec{k}/\sqrt{3\mu E}$ and $q\equiv \ltvec{q'}/\sqrt{3\mu E}$ the integral to consider is
\begin{align}
    \label{eq:smallshiftlargex}
    \langle\ltvec{p}^2\rangle_{I2} & \equiv 2\pi\mu^2\int_{\alpha'}^1dx\int_0^1dk\int_0^{2\pi}d\theta\int_0^{q_+(k,\theta)}dq \nonumber\\ & \quad\; \frac{1}{x}\frac{k-q\cos\theta}{(k^2+q^2-2kq\cos\theta+\alpha^2)^2}\frac{q^5}{x^2\beta^2+q^4}, \\
    q_+ & \equiv k\cos\theta+\sqrt{1-k^2\sin^2\theta},
\end{align}
where $\alpha\equiv\sqrt{\mu/3E}$, $\beta\equiv 4/3\mu L$, and $\alpha'\equiv 3\alpha/2$.  (For notational simplicity we've dropped the overall factors of \linebreak $C_R\alpha_s L/\pi^2\lambda$.)  GLV energy loss is derived in the limit $L\gg1/\mu$, so we will always have $\beta\ll1$.  The large energy limit corresponds to taking $\alpha\ll1$.  One may straightforwardly perform the $x$ integration to yield
\begin{multline}
    \langle\ltvec{p}^2\rangle_{I2} = 2\pi\mu^2\int_0^1dk\int_0^{2\pi}d\theta\int_0^{q_+(k,\theta)}dq \nonumber\\ \frac{k-q\cos\theta}{(k^2+q^2-2kq\cos\theta+\alpha^2)^2}\frac{1}{2}q\ln\Big(\frac{q^4+\alpha^{\prime2}\beta^2}{(q^4+\beta^2)\alpha^{\prime2}}\Big),
\end{multline}
The strong dropoff of the original integrand with $\sim\ltvec{q}^{-2}$ suggests that the result may be insensitive to the $q$ upper bound; numerically, one finds small and decreasing corrections to the exact result when one taken the upper bound of the $q$ integral to infinity, $q_+\rightarrow\infinity$.  Taking $q_+\rightarrow\infinity$ we are able to immediately perform both the $k$ and $\theta$ integrals, yielding
\begin{multline}
    \langle\ltvec{p}^2\rangle_{I2} \simeq -2\pi\mu^2\int_0^\infinity dq \frac{\pi q}{2}\\ \frac{ q^2+\alpha^2-\sqrt{q^4-2q^2(1-\alpha^2)+(1+\alpha^2)^2}}{(q^2+\alpha^2)\sqrt{q^4-2q^2(1-\alpha^2)+(1+\alpha^2)^2}} \\ \ln\Big(\frac{q^4+\alpha^{\prime2}\beta^2}{(q^4+\beta^2)\alpha^{\prime2}}\Big),
\end{multline}
where we explicitly note with the $\simeq$ the approximation made by taking the upper limit of the $q$ integration to infinity.  Of the result, notice first the emergence of an overall minus sign.  Second, notice the explicit delicate subtraction occurring in the numerator of the ratio in the integrand.  We separately evaluate the contributions from the two subtracting terms.  Each contribution individually UV log diverges, so we must artificially cut off the upper limit of the integral; we'll call this artificial cutoff $u$.  We will take $u\rightarrow\infinity$ when we put the two contributions together.

One may evaluate
\begin{align}
    \int_0^{u^2} dq^2 \frac{1}{q^2+\alpha^2} \ln\Big(\frac{q^4+\alpha^{\prime2}\beta^2}{(q^4+\beta^2)\alpha^{\prime2}}\Big)
\end{align}
in closed form in an uninsightful combination of logs and dilogarithms.  

\begin{figure}
    \centering
    \includegraphics[width=0.9\columnwidth]{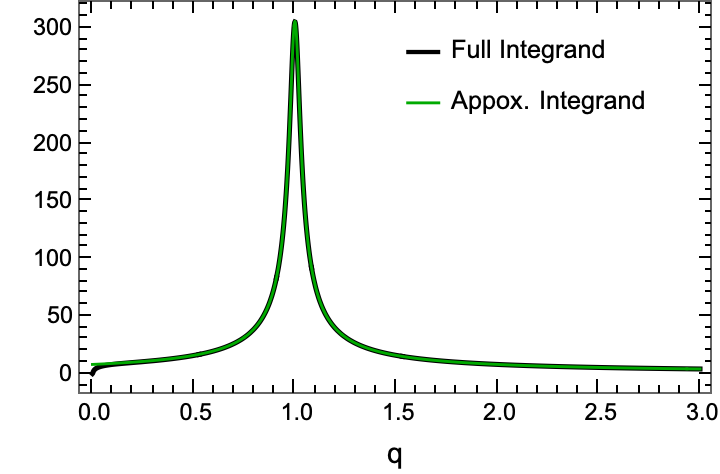}
    \caption{Comparison between the full integrand \protect\eq{eq:fullintegrand} (thick, black) and the approximate integrand \eq{eq:approxintegrand} (thin, green) for $E=1000$ GeV, $L=5$ fm and $\mu=0.5$ GeV.}
    \label{fig:approxint}
\end{figure}

The integral
\begin{multline}
    \label{eq:fullintegrand}
    \int_0^{u^2} dq^2 \frac{\ln\Big(\frac{q^4+\alpha^{\prime2}\beta^2}{(q^4+\beta^2)\alpha^{\prime2}}\Big)}{\sqrt{q^4-2q^2(1-\alpha^2)+(1+\alpha^2)^2}} 
\end{multline}
cannot be evaluated in closed form.  However, the integrand is highly peaked around $q=1$, and the integral is dominated by the region around $q=1$.  When $q\simeq1$, $\beta\ll1$ and $\alpha\ll1$ implies that the argument of the log can be approximated by
\begin{align}
    \frac{q^4+\alpha^{\prime2}\beta^2}{(q^4+\beta^2)\alpha^{\prime2}} \simeq \frac{1+\alpha^{\prime2}\beta^2}{(1+\beta^2)\alpha^{\prime2}} \approx \frac{1}{\alpha^{\prime2}}.
\end{align}
This is a particularly good approximation for the entire integration region: the log only serves to enhance the dying off of the integrand for both large and small $q$.  But for large $q$ the integrand is already dying off like $1/q^2$, and for small $q$ the integral is already dying off like $q$.  See \fig{fig:approxint} to see just how good the approximation is for $E=1000$ GeV, $L=5$ fm and $\mu=0.5$ GeV.

For large $u^2$ we have
\begin{multline}
    \label{eq:approxintegrand}
    \int_0^{u^2} dq^2 \frac{\ln(1/\alpha^{\prime2})}{\sqrt{q^4-2q^2(1-\alpha^2)+(1+\alpha^2)^2}} \\ \approx \ln(1/\alpha^{\prime2})\ln(u^2/\alpha^2).  
\end{multline}

\begin{figure}
    \centering
    \includegraphics[width=0.9\columnwidth]{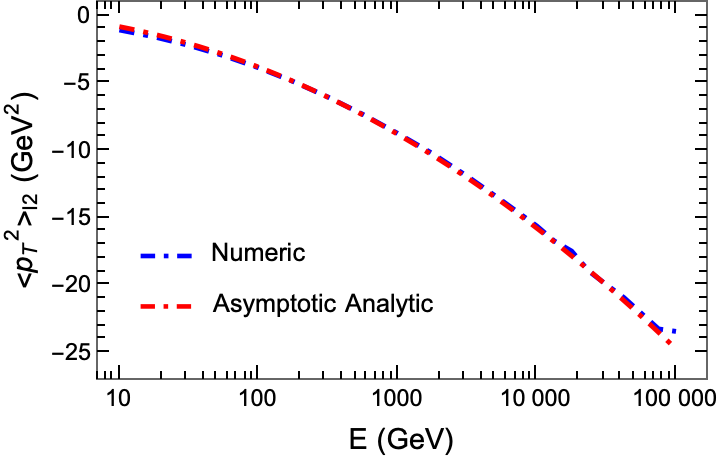}
    \caption{Comparison between the full numerical $\langle\ltvec{p}^2\rangle_{I2}$ from \protect\eq{eq:smallshiftlargex} to the approximation \protect\eq{eq:smallshiftlargexapprox} for $L=5$ fm and $\mu=0.5$ GeV with the overall factor of $C_R\alpha_s L/\pi^2\lambda$ restored.}
    \label{fig:smallshiftlargexcomparison}
\end{figure}

When the two contributions are combined, the $\ln(u^2)$ divergences cancel (as they must) and the remainder is
\begin{multline}
    \label{eq:smallshiftlargexapprox}
    \langle\ltvec{p}^2\rangle_{I2} = {-}\frac12\pi^2\mu^2\ln(\frac32\alpha)\Big(3\ln(\alpha) \\ - 2\ln(\beta) - \ln(\frac32) \Big). 
\end{multline}
We compare the full numerical $\langle\ltvec{p}^2\rangle_{I2}$ from \eq{eq:smallshiftlargex} to the approximation \eq{eq:smallshiftlargexapprox} (with the overall factor of \linebreak $C_R\alpha_s L/\pi^2\lambda$ restored) in \fig{fig:smallshiftlargexcomparison}.

\subsubsection{Small Shift, Small \texorpdfstring{$x$}{x} Integral}
As was done for the small shift, large $x$ integral, let's first make the integral dimensionless by scaling out $\sqrt{3\mu E}$ from both $\ltvec{k}$ and $\ltvec{q'}$.  Defining $k\equiv \ltvec{k}/\sqrt{3\mu E}$ and $q\equiv \ltvec{q'}/\sqrt{3\mu E}$ the integral to consider is
\begin{multline}
    \label{eq:smallshiftsmallx}
    \langle\ltvec{p}^2\rangle_{I1} \equiv 2\pi\mu^2\int_0^{\alpha'}dx\int_0^{x/\alpha'}dk\int_0^{2\pi}d\theta\int_0^{q_+(k,\theta)}dq \\ \frac{1}{x}\frac{k-q\cos\theta}{(k^2+q^2-2kq\cos\theta+\alpha^2)^2}\frac{q^5}{x^2\beta^2+q^4},
\end{multline}

From the intuition gained from the small shift, large $x$ integral, one can check numerically that lifting the upper bound of the $q$ integral to infinity, $q_+\rightarrow\infinity$, is a negligible change.  With the dependence on $k$ and $\theta$ removed from the upper limit of $q$ we may perform the $\theta$ and $k$ integrals.  We may further define $y\equiv (x/\alpha')^2$.  Then we are left with
\begin{multline}
    \langle\ltvec{p}^2\rangle_{I1} \simeq 2\pi\mu^2\int_0^{1}dy\int_0^{\infinity}dq \frac{\pi}{2y}\Big( \frac{1}{q^2+\alpha^2} \\ - \frac{1}{\sqrt{y^2-2y(q^2-\alpha^2)+(q^2+\alpha^2)^2}} \Big)\frac{q^5}{y\beta^{\prime2}+q^4},
\end{multline}  
where $\beta'\equiv\alpha'\beta$.  We see again a  subtraction leading to a delicate cancellation of (this time IR) divergences.  

We may readily perform the $y$ integral over the integrand resulting from the first term in the parentheses.  Temporarily inserting an IR regulator $\epsilon$ to be taken to 0, one has that
\begin{multline}
    \int_\epsilon^1dy\frac{q^5}{y(q^2+\alpha^2)(y\beta'+q^4)} \\ = \frac{q}{(q^2+\alpha^2)}\ln\big( \frac{q^4}{(\beta'+q^4)\epsilon} \big).
\end{multline}

One may also perform the integral over $y$ from the integrand resulting from the second term in the parentheses.  The result is complicated, including powers of $q$ and $\alpha$, square roots, and arctanh's of complicated arguments.  When expanding the result for small $\epsilon$ one finds exactly the $\ln\epsilon$ required to cancel the IR divergence from the first term.  Numerically, it turns out the remaining contributions delicately approximately cancel each other and yield a negligible contribution in the high energy limit.

We are thus left to compute
\begin{align}
    \int_0^{\infinity} dq^2\frac{1}{q^2+\alpha^2}\ln\big( \frac{q^4}{\beta^{\prime2}+q^4} \big),
\end{align}
which yields another uninsightful combination of logs and dilogarithms.  

However, in the limit of very small $\alpha$ one finds that
\begin{multline}
    \label{eq:smallshiftsmallxapprox}
    \langle\ltvec{p}^2\rangle_{I1} = -\frac{1}{24}\pi^2\mu^2\big( 12\ln^2\alpha-24\ln(\frac32\beta)\ln\alpha \\ +12\ln^2(\frac32\beta)+5\pi^2 \big). 
\end{multline}

\begin{figure}
    \centering
    \includegraphics[width=0.9\columnwidth]{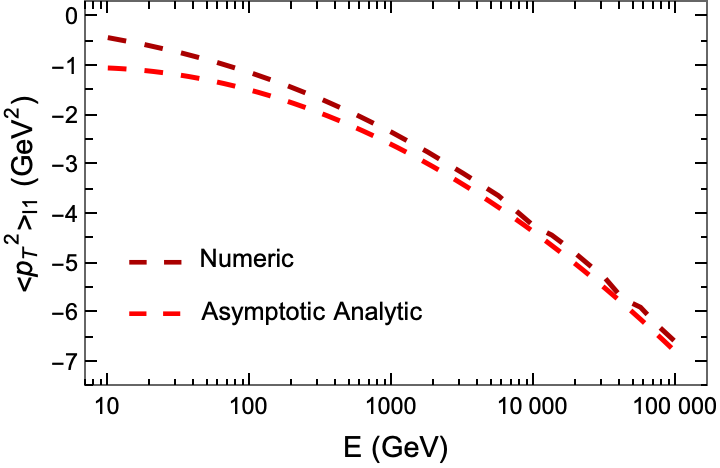}
    \caption{Comparison between the full numerical $\langle\ltvec{p}^2\rangle_{I1}$ from \protect\eq{eq:smallshiftsmallx} to the approximation \protect\eq{eq:smallshiftsmallxapprox} for $L=5$ fm and $\mu=0.5$ GeV with the overall factor of $C_R\alpha_s L/\pi^2\lambda$ restored.}
    \label{fig:smallshiftsmallxcomparison}
\end{figure}

We show in \fig{fig:smallshiftsmallxcomparison} a comparison between the full numerical $\langle\ltvec{p}^2\rangle_{I1}$ from \eq{eq:smallshiftsmallx} to the approximation \eq{eq:smallshiftsmallxapprox} for $L=5$ fm and $\mu=0.5$ GeV and with the overall factor of $C_R\alpha_s L/\pi^2\lambda$ restored.

\subsubsection{Complete Leading Order in \texorpdfstring{$E$}{E} Expansion}
\label{sec:asymptoticradiative}
Combining the leading order in energy results from the previous two subsubsections leads to a surprising simplification.  Restoring the relevant prefactors, we find
\begin{align}
    \label{eq:leadingE}
    \langle\ltvec{p}^2\rangle_{\text{NLO, }1} = -\frac{C_R\alpha_s}{4}\frac{L}{\lambda}\mu^2\left[ \ln^2\Big( \frac{4E}{\mu^2L} \Big) + \frac{5\pi^2}{12} \right]. 
\end{align}

\begin{figure}
    \centering
    \includegraphics[width=0.9\columnwidth]{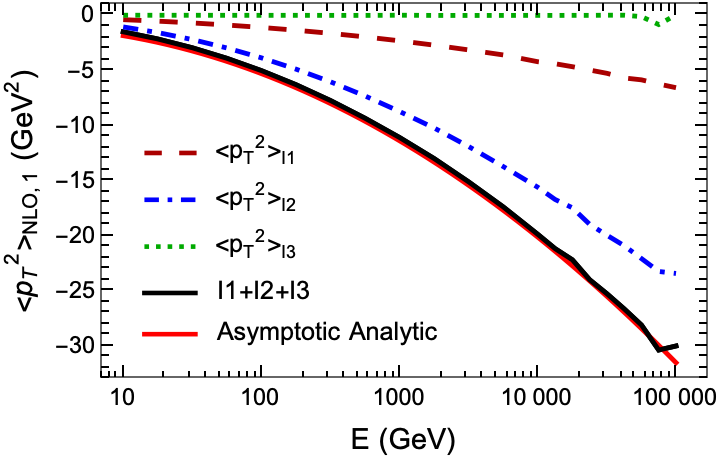}\\[0.2in]
    \includegraphics[width=0.9\columnwidth]{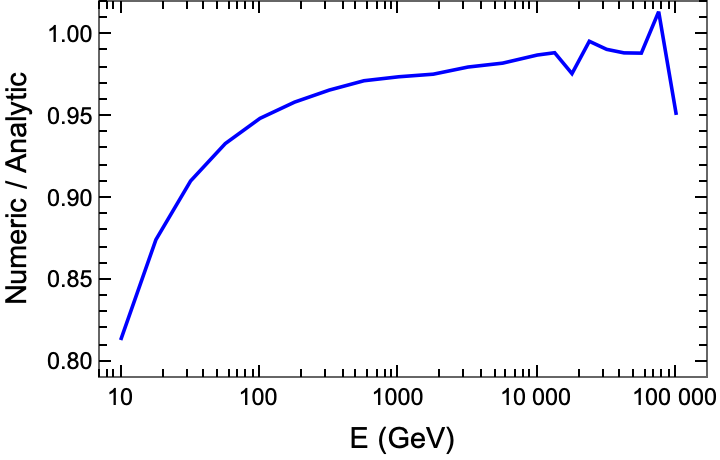}
    \caption{(Top) Comparison between the full result for the momentum broadening from radiative emissions from \protect\eq{eqn:pT2NLO} (dots) and the leading energy behavior from \protect\eq{eq:leadingE} for $L=5$ fm and $\mu=0.5$ GeV.  (Bottom) Ratio of the full numerical result \protect\eq{eqn:pT2NLO} to the asymptotic analytic result \protect\eq{eq:leadingE} for $L=5$ fm and $\mu=0.5$ GeV.}
    \label{fig:approx}
\end{figure}

We show in \fig{fig:approx} (top) a comparison of the leading in energy approximation from \eq{eq:leadingE} with the full result from \eq{eqn:pT2NLO} for $L=5$ fm and $\mu = 0.5$ GeV.  In \fig{fig:approx} (bottom) we show the ratio of the numeric to analytic expressions.  One can see especially well in the ratio plot the onset of significant numerical instabilities for $E\gtrsim2000$ GeV.  Despite the significant numerical instabilities, the analytic approximation appears to do an extremely good job of capturing the large energy behavior of the full numerical result.

With the approximations detailed in the previous subsubsections, \eq{eq:leadingE} is correct to $\mathcal O(E^{-1})$.  One should question the confidence we have in the lack of a subleading log and also our confidence in the constant in \eq{eq:leadingE}.  While none of our calculations were performed with absolute mathematical rigor---we did not rigorously assess the importance of the approximations we made---the systematic comparison with numerical results gives us a very high degree of confidence in the lack of any significant subleading log.  We did completely neglect the nearly energy-independent contribution from the large shift, $I3$.  Whether this contribution from $I3$ grows with a log or is approximately energy independent isn't completely clear; this integral is the least numerically stable of all three.  Nevertheless, the large shift $I3$ is orders of magnitude smaller than the small shift $I1$ and $I2$.  We thus believe our result \eq{eq:leadingE} to a very good approximation fully holds to $\mathcal O(E^{-1})$.

In order to more readily compare with other approaches, one can also express \eq{eq:leadingE} as a ratio of the radiative to collisional broadening components.  Dividing \eqref{eq:leadingE} by \eqref{e:CollAsym} gives
\begin{align}   \label{eq:leadingEratio}
    \frac{\langle p_T^2 \rangle_\theRIGHTwayNLOone}
    {\langle p_T^2 \rangle_\theRIGHTwayLOone} =
    - \frac{\alpha_s C_R}{4} \ln\frac{E}{\mu} =
    - \frac{\alpha_s}{3} \ln\frac{E}{\mu}. 
\end{align}
One can see that this ratio of the correct, finite kinematics asymptotic radiative energy loss with the elastic energy loss differs from the same ratio as computed with the infinite kinematics given by \eqref{e:DGLVratio} by a factor of $- \pi / 4 \approx -0.79$.

%
\section{Discussion of Results}
\label{sec:discussion}
%

%
\subsection{Comparison to Kang et al.}
%
Now that we have the asymptotic behavior of the radiative energy loss contribution to jet broadening under control, we'd like to interpret our results.  In particular, we'd like to compare the energy loss approach, which assumes a factorization of the production process from the subsequent in-me\-di\-um evolution, and the twist-4 approach of \cite{Kang:2013raa,Xing:2014kpa,Kang:2014ela}, which does not assume \textit{a priori} such a factorization.

We'll first compare at leading order, which in the energy loss formalism is elastic energy loss.
\subsubsection{Leading Order}
\label{sec:LOcomparison}
From \cite{Kang:2014ela}, the leading order in $\alpha_s$ twist-4 contribution to jet broadening is
\begin{multline}
    \label{eq:tw4el}
    \frac{d\langle\ltvec{\ell}^2\sigma\rangle}{dx_Bdydz_h} = \sigma_h e_q^2 \int_{x_B}^1\frac{dx}{x}T_{qg}(x,0,0,\mu_f^2) \\ \int_{z_h}^1\frac{dz}{z}D_{h/q}(z,\mu_f^2)\delta(1-\hat x)\delta(1-\hat z),
\end{multline}
where
\begin{align}
    \sigma_h & = \frac{4\pi^2\alpha_sz_h^2}{N_c}\sigma_0 \\
    \sigma_0 & = \frac{2\pi\alpha_{EM}^2}{Q^2}\frac{1+(1-y)^2}{y} \\
    \hat x & = \frac{x_B}{x};\quad \hat z = \frac{z_h}{z}.
\end{align}
$T_{qg}$ is the twist-4 quark-gluon correlation function, a generalization of the usual twist-2 parton distribution function.  In the limit of a large and loosely bound nucleus, in which one may neglect the spatial and momentum correlations between the two nucleons, one has \cite{Kang:2014ela} an approximate factorization 
\begin{align}
    T_{qg}(x_B,0,\mu_f^2) 
    & \approx \frac{N_c}{4\pi^2\alpha_s}f_{q/A}(x_B,\mu_f^2)\int dy^-\hat q(\mu_f^2,y^-) \nonumber\\
    \label{eq:tw4approx}
    & = \frac{N_c}{4\pi^2\alpha_s}f_{q/A}(x_B,\mu_f^2)\hat q(\mu_f^2)L,
\end{align}
where in the last line we assumed for simplicity that the parton propagates through a nucleus of constant density of thickness $L$.

In order to most readily and clearly compare to the energy loss derivation that we will show below, we will remove the complication of the fragmentation process from the twist-4 approach by assuming exact parton-hadron duality, i.e.\ we will take
\begin{align}
    D_{h/q}(z,\mu_f^2) = \delta(1-z).
\end{align}
We then have that
\begin{align}
    \label{eq:tw4frag}
    \int_0^1dz_h\int_{z_h}^1\frac{dz}{z}D_{h/q}(z,\mu_f^2)\delta(1-\hat z) = 1.
\end{align}

Putting together Eqs.\ (\ref{eq:tw4el}), (\ref{eq:tw4approx}), and (\ref{eq:tw4frag}), we have for the leading in $\alpha_s$ contribution from the twist-4 approach a completely factorized result
\begin{align}
    \label{eq:tw4asympel}
    \frac{d\langle\ltvec{\ell}^2\sigma\rangle}{dx_Bdy} = \frac{d\sigma_0}{dx_Bdy}\hat q(\mu_f^2)L,
\end{align}
where
\begin{align}
    \frac{d\sigma_0}{dx_Bdy} \equiv \frac{4\pi\alpha_{EM}^2e_q^2}{yQ^2}(1-y+\frac12y^2)f_{q/A}(x_B,\mu_f^2)
\end{align}
gives the differential production cross section.

On the other hand, as was shown in \secref{sec:LO1st}, a simple estimate for the jet broadening from $2\rightarrow2$ elastic scattering in medium is given by
\begin{align}
    \langle p_T^2 \rangle_{\text{LO, }1} 
    &\approx
    \frac{L\mu^2}{\lambda} \ln(\frac{E}{\mu}) \: .
\end{align}
If we identify a ``running'' $\hat q$ from the energy loss perspective 
\begin{align}
    \hat q(\mu) \equiv \frac{\mu^2}{\lambda}\ln\big( \frac{E}{\mu} \big),
\end{align}
as was done in \secref{sec:LO1st}, then we see an exact equivalence between the leading order in $\alpha_s$ result from twist-4, \eq{eq:tw4asympel}, and the energy loss result \eq{e:CollAsym}.  (Recall that in the energy loss approach the in-medium jet broadening is \textit{conditional} on the production of a high momentum parton; hence the production cross section $d\sigma_0/dx_Bdy$ is divided out.)

\subsubsection{Next-to-Leading Order}
\label{sec:NLOcomparison}
We would now like to go one step further, building on the asymptotic analysis of the previous section, and attempt to compare the asymptotics of the radiative energy loss contribution to jet broadening to the asymptotics of the next-to-leading order contribution to jet broadening from the twist-4 approach.

As a first step, we would like to compare the leading asymptotics of the two approaches.  We saw in \secref{sec:LOcomparison} that the leading order energy loss contribution to jet broadening grows with the log of energy.  As we saw in \secref{sec:asymptoticradiative}, the leading order asymptotics of the jet broadening from radiative emissions within the energy loss approach shows a leading double logarithmic growth with energy.  The full twist-4 next-to-leading order result from \cite{Kang:2013raa,Xing:2014kpa,Kang:2014ela} includes both leading as well as subleading contributions in energy.  We will focus here only on the leading contribution.  From \cite{Kang:2014ela} the twist-4 approach has an overall log enhanced contribution given by\\
\begin{widetext}
\begin{multline}
    \frac{d\langle \ltvec{\ell}^2\sigma\rangle}{dx_Bdydz_h}\bigg\rvert_{\text{NLO}} = \sigma_h\frac{\alpha_s}{2\pi} e_q^2 \ln\Big(\frac{Q^2}{\mu_f^2}\Big) \int_{z_h}^1\frac{dz}{z}D_{h/q}(z,\mu_f^2)\int_{x_B}^1\frac{dx}{x}\Big\{ \\ \delta(1-\hat x)P_{qq}(\hat z)T_{qg}(x,0,0,\mu_f^2)+\delta(1-\hat z)\big(P_{qg\rightarrow qg}\otimes T_{qg}+P_{qg}(\hat x)T_{gg}(x,0,0,\mu_f^2)\big)\Big\},
\end{multline}
\begin{multline}
    P_{qg\rightarrow qg}\otimes T_{qg} \equiv P_{qq}(\hat x) T_{qg}(x,0,0)+\frac{C_A}{2}\bigg\{ \frac{4}{(1-\hat x)_+}T_{qg}(x_B,x-x_B,0)-\frac{1+\hat x}{(1-\hat x)_+}\big[ T_{qg}(x,0,x_B-x) \\ +T_{qg}(x_B,x-x_B,x-x_B) \big]  \bigg\} +2C_A\delta(1-\hat x)T_{qg}(x,0,0).
\end{multline}
\end{widetext}
\indent We would again like to isolate and trivialize the fragmentation function contribution to ease the comparison to the energy loss approach.  At leading order, we could accomplish this trivialization by simply replacing the fragmentation function with a delta function.  At next-to-leading order, trivializing the fragmentation function contribution is more difficult as we must remove not only the fragmentation function but also its evolution.  The term proportional to $\delta(1-\hat x)$ is exactly this NLO evolution of the fragmentation function.  Removing this evolution and replacing $D_{h/q}(z,\mu_f^2)\rightarrow\delta(1-z)$ we are left with
\vspace{-.1in}
\begin{multline}
    \frac{d\langle \ltvec{\ell}^2\sigma\rangle}{dx_Bdy}\bigg\rvert_{\text{NLO}} = \sigma_h\frac{\alpha_s}{2\pi} e_q^2 \ln\Big(\frac{Q^2}{\mu_f^2}\Big)\int_{x_B}^1\frac{dx}{x}\Big\{ \\ P_{qg\rightarrow qg}\otimes T_{qg}+P_{qg}(\hat x)T_{gg}(x,0,0,\mu_f^2) \Big\}.
\end{multline}

One immediately sees that---unlike the energy loss ap\-proach---the twist-4 approach involves significantly more physics.  The result is clearly color non-trivial: there are several contributions proportional to $C_A$.  There is also a mixing of the $T_{qg}$ and $T_{gg}$ twist-4 distribution functions.  

If we assume that the color triviality breaking terms are small compared to the $C_F$ behavior, then we have that
\begin{multline}    \label{e:ZBKasym}
    \frac{d\langle \ltvec{\ell}^2\sigma\rangle}{dx_Bdy}\bigg\rvert_{\text{NLO}} 
     = \sigma_h\frac{\alpha_s}{2\pi} e_q^2 \ln\Big(\frac{Q^2}{\mu_f^2}\Big)\int_{x_B}^1\frac{dx}{x}\bigg[  \\ 
     \qquad P_{qq}(\hat x)T_{qg}(x,0,0,\mu_f^2) +P_{qg}(\hat x)T_{gg}(x,0,0,\mu_f^2) \bigg].
\end{multline}

If we again assume a large and loosely bound nucleus, we again can take that the twist-4 distribution functions factorize.  For $T_{qg}$ we have again \eq{eq:tw4approx}.  For $T_{gg}$ we have the obvious generalization
\begin{align}
    T_{gg}(x_B,0,\mu_f^2) 
    & \approx \frac{N_c}{4\pi^2\alpha_s}f_{g/A}(x_B,\mu_f^2)\int dy^-\hat q(\mu_f^2,y^-) \nonumber\\
    & = \frac{N_c}{4\pi^2\alpha_s}f_{g/A}(x_B,\mu_f^2)\hat q(\mu_f^2)L.
\end{align}
Note that the $\hat q$ in the above is the same as in \eq{eq:tw4approx}: in both cases it is the high-momentum quark that is propagating through and being kicked by the nucleus.  

We therefore find that at next-to-leading order in the twist-4 approach, assuming small color triviality violating contributions, the jet broadening is given by
\begin{align}
    \label{eq:NLOtw4}
    \frac{d\langle \ltvec{\ell}^2\sigma\rangle}{dx_Bdy}\bigg\rvert_{\text{NLO}} 
     = \frac{d\sigma_0}{dx_Bdy}_{LL}\hat q(\mu_f^2)L,
\end{align}
where $d\sigma_0/dx_Bdy|_{LL}$ is the leading logarithmic, next-to-lead\-ing order production cross section for a hard parton in a DIS event \cite{Paukkunen:2009ks}.  

In order to facilitate comparison between the twist-4 result and the energy loss result, we consider the ratio of the radiative (NLO) component \eqref{e:ZBKasym} and collisional (LO) component \eqref{eq:tw4asympel} within the twist-4 formalism.  Note also that the observable $d\langle \ell_\bot^2 \sigma \rangle / dx_B dy$ is proportional to $\langle p_T^2 \rangle$ up to a normalization factor which cancels in the ratio.  Thus we may write for the twist-4 formalism
\begin{align}   \label{e:ZBKratio}
    &\frac{\langle p_T^2 \rangle_\theRIGHTwayNLOone}{\langle p_T^2 \rangle_\theRIGHTwayLOone} = 
    \frac{\alpha_s}{2\pi} \ln\frac{Q^2}{\mu_f^2} 
    \notag \\ &\times
    \frac{
        \int_{x_B}^1 \frac{dx}{x} \left[
        P_{q q} (\hat{x}) f_{q/A} (x, \mu_f^2) +
        P_{q g} (\hat{x}) f_{g/A} (x, \mu_f^2)
        \right]
    }{
        f_{q/A} (x_B, \mu_f^2)
    }
    \notag \\ \notag \\ &\approx
    \frac{4 \alpha_s}{3\pi} \ln\frac{E}{\mu} 
    \notag \\ & \hspace{0.5cm} \times
    \frac{
        \int_{x_B}^1 \frac{dx}{x} \left[
        \frac{1 + \hat{x}^2}{(1 - \hat{x}_+} + \frac{3}{2} \delta(1-\hat{x})
        \right] f_{q/A} (x, \mu^2)
    }{
        f_{q/A} (x_B, \mu^2)
    } \: ,
\end{align}
where the last line follows for a target composed of elementary quarks $f_{g / A} \rightarrow 0$, setting the factorization scale $\mu_f = \mu$, and identifying the hard scale $Q$ as the jet energy $E$.

There are very important remarks to make when we compare the leading twist-4 broadening  \eq{eq:NLOtw4} with the leading broadening from the energy loss approach \eq{eq:leadingE}.  First, if we again interpret $\hat q(\mu) \sim (\mu^2/\lambda)\ln(E/\mu)$, then we again see agreement in terms of the strength of the growth: both the twist-4 and energy loss broadenings grow like $\ln^2(E)$.  However, we learn something very interesting about the energy loss approach by comparing to the twist-4 approach: the twist-4 approach tells us that the leading logarithmic growth in $\langle p_T^2 \rangle$ as computed in the energy loss approach should really be associated with a modification of the initial state parton distribution function, rather than a modification of the fragmentation function.   
This point is further emphasized by the ratio \eqref{e:ZBKratio} of the radiative to collisional broadening components in the twist-4 framework.  Clearly the ratio of radiative to collisional broadening is linked in the twist-4 formalism to the $x$ dependence of the PDFs -- a feature that is notably absent from the energy loss framework.  Moreover, if one neglects the modification of the PDFs between LO and NLO by hand, setting the ratio of PDFs in the second line of \eq{e:ZBKratio} to unity, then the coefficient $\frac{4 \alpha_s}{3\pi} \ln\tfrac{E}{\mu}$ agrees \textit{exactly} with the one obtained in \eq{e:DGLVratio} for the DGLV formalism under the (oversimplified) assumption of infinite kinematics.  This prediction of broadening, however, puts the twist-4 prediction in stark contrast with the DGLV high-energy asymptotics obtained with exact kinematics in Sec.~\ref{sec:finitebroad}: although the leading logarithmic energy dependence is the same between the two approaches, the coefficients in the two approaches have the opposite signs.  While the true prediction of the DGLV formalism with finite kinematic bounds is a net narrowing of the transverse momentum distribution, the twist-4 approach predicts a broadening.  

Our careful analysis of the effect of the kinematic limits suggests that the broadening predicted by the twist-4 approach is an artifact of neglecting kinematics limits as is done in the usual collinear factorization approach: integrating over all $\ltvec{k}$ of the emitted gluon is a bad approximation to the correct limited kinematics and leads to the wrong sign for the coefficient of the leading double logarithmic contribution to jet broadening.

With the above said, one may naturally ask: what do the data show?

%
\subsection{Comparison to LHC Data}
\label{sec:LHCcomparison}
%

Given the range of predictions for the transverse momentum broadening seen numerically and analytically, under different approximations to the kinematic limits, it is prudent to look to experiment to benchmark our expectations.

The ALICE collaboration presented a new approach \cite{ALICE:2015mdb} to the measurement of jet quenching, which was based on the semi-inclusive distribution of charged jets recoiling from a high transverse momentum charged hadron trigger in 0-10\% central Pb-Pb collisions at $\sqrt{s_{NN}} = 2.76$ TeV. The collaboration also investigated the medium-induced a\-co\-plan\-ar\-i\-ty, or ``inter-jet broadening," by extending their analysis to the measurement of the angular distribution of recoil jet yield with respect to the axis defined by the trigger hadron momentum. The azimuthal correlation between the trigger hadron and coincident recoil charged jets is measured via the distribution $\Phi(\Delta\varphi)$.

\begin{figure}
    \centering
    \includegraphics[width=0.9\columnwidth]{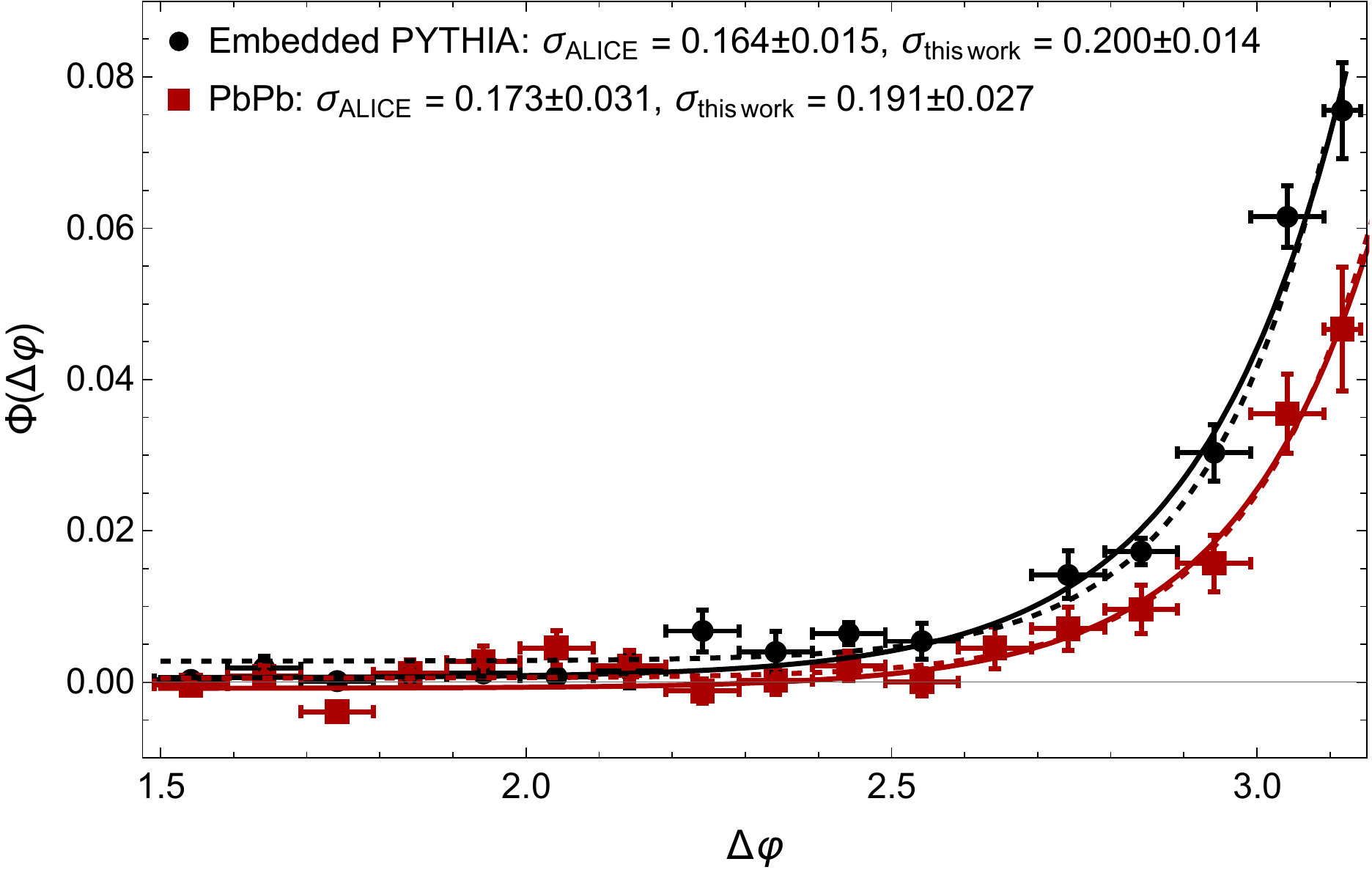}
    \caption{The $\Phi(\Delta\varphi)$ distributions for 0-10\% centrality Pb-Pb collisions (red) measured by the ALICE collaboration \cite{ALICE:2015mdb} and PYTHIA-simulated pp collisions embedded into central Pb-Pb collisions (black), at $\sqrt{s_{NN}} = 2.76$ TeV. The error bars indicate statistical errors only. Both datasets are fit to exponential functions of the form shown in \eqref{eqn:ALICEfit}; this work fits over the whole $\Delta\varphi$ range (solid lines), while the ALICE collaboration fitted over $2\pi/3 < \Delta\varphi < \pi$ (dashed lines). Both sets of fits are shown by the lower red (Pb-Pb) and upper black (pp) lines. }\label{fig:ALICEdata}
\end{figure}

Figure~\ref{fig:ALICEdata} shows the $\Phi(\Delta\varphi)$ distribution for 0-10\% central Pb-Pb collisions at $\sqrt{s_{NN}} = 2.76$ TeV measured by the ALICE collaboration \cite{ALICE:2015mdb}. This data includes jets with $40<p_{\text{T,jet}}^{\text{reco,ch}}<60$ GeV/c, where the reconstructed charged jet transverse momentum $p_{\text{T,jet}}^{\text{reco,ch}}$ is not corrected for background fluctuations and instrumental effects. Due to the insufficient statistical significance of the current data for pp collisions at $\sqrt{s} = 2.76$ TeV, the reference pp distribution for the reported Pb-Pb measurements was calculated using the PYTHIA event generator. The simulated reference distributions were validated through comparison with ALICE data of pp collisions at $\sqrt{s} = 7$ TeV. The pp data points shown in Figure~\ref{fig:ALICEdata} were simulated using PYTHIA 6.425 with the Perugia 2010 tune, and are modified by the expected instrumental and background effects for central Pb-Pb collisions. 

The $\Phi(\Delta\varphi)$ distributions for Pb-Pb and pp collisions are fit to functions of the form
\begin{equation}\label{eqn:ALICEfit}
    f(\Delta\varphi) = p_0 \times e^{(\Delta\varphi-\pi)/\sigma} + p_1 \, ,
\end{equation}
where the width of the exponential distribution is characterized by the parameter $\sigma$. 

The ALICE collaboration fit the pp and Pb-Pb data shown in Figure~\ref{fig:ALICEdata} to the function \eqref{eqn:ALICEfit} over the range $2\pi/3 < \Delta\varphi < \pi$ and obtained the width fit parameters $\sigma_{\text{Pb-Pb}}=0.173$ $\pm 0.031 \text{(stat.)} \pm 0.005 \text{(sys.)}$ and $\sigma_{\text{PYTHIA}}=0.164\pm0.015$\linebreak $\text{(stat.)}$.  Nominally, these results indicate a broadening of the acoplanarity distribution due to the medium: $\sigma_{\text{Pb-Pb}}>$ \linebreak $\sigma_{\text{PYTHIA}}$.  However, the two cases are also consistent within uncertainties ($\sigma_{\text{Pb-Pb}} = \sigma_{\text{PYTHIA}}$).  On the other hand, we can compare this with a fit we have performed over the entire measured $\Delta\varphi$ region, obtaining $\sigma_{\text{PYTHIA}}=0.200\pm0.014$ and $\sigma_{\text{Pb-Pb}}=0.191\pm0.027$. These fit parameters also agree within uncertainties, but now nominally suggest a slight narrowing ($\sigma_{\text{Pb-Pb}}<\sigma_{\text{PYTHIA}}$).

The ALICE data presented here are inconclusive about the modification of the jet broadening distribution due to the medium.  Depending on the region over which one performs the exponential fit, one can infer either a narrowing or a broadening, and overall neither set of results provide conclusive evidence for any medium-induced acoplanarity of recoil jets.  Clearly, differentiating experimentally between different theoretical formalisms is challenging and requires further development both theoretically and experimentally.

%
\section{Discussion and Conclusions}
\label{sec:concl}
%

In this work, we studied jet-medium interactions at first order in opacity in the DGLV energy loss formalism \cite{Gyulassy:2000er,Djordjevic:2003zk} and compared it with the predictions of the collinear twist-4 formalism of Kang et al.\ \cite{Kang:2014ela,Xing:2014kpa, Kang:2013raa}.  We find that the opacity expansion predicts a leading order (collisional) momentum broadening that grows like $\ln E/\mu$ with a next-to-leading order (radiative) momentum \textit{narrowing} due to destructive LPM effects, a narrowing that grows like $\ln^2E/\mu$.  We find that the leading order opacity expansion broadening agrees exactly with the leading order twist-4 broadening.  The next-to-leading order twist-4 broadening includes many terms absent from the opacity expansion next-to-leading order asymptotics (including terms that break color triviality and that are subleading in energy).  Most important, though, the twist expansion appears to predict a jet broadening (as opposed to the narrowing from the opacity expansion).  This qualitative difference is likely due to a less careful treatment of finite kinematics and the LPM effect in the twist-4 approach.  At the same time, the twist-4 approach shows that ``final state'' energy loss manifests in the modification of initial state objects that are the natural twist-4 generalization of the twist-2 parton distribution functions.  This in striking contrast to their interpretation as ``final state effects'' in energy loss approaches, such as their use as kernels of ``medium-modified DGLAP evolution'' \cite{Guo:2000nz,Wang:2001ifa,Armesto:2007dt,Aurenche:2008hm,Majumder:2009ge,Chien:2015vja,Sirimanna:2021sqx}.

To discriminate between the predictions of these two formalisms, the medium-modification effects must be distinguishable from the background: energy loss and transverse momentum broadening which occurs already in vacuum.  As summarized in Eqs.~\eqref{e:vacasym}, the number of emitted gluons and the fractional energy loss both grow logarithmically (as powers of $\ln\tfrac{E}{\mu}$).  However the mean momentum broadening $\langle p_T^2 \rangle_\theRIGHTwayNLOzero$ grows much faster -- quadratically with $E$ -- due to the $k_\bot^2$ weighting of the splitting function.  The strong dominance of this vacuum broadening is clearly seen from the numerical calculation of Sec.~\ref{sec:numerics} as illustrated in Fig.~\ref{fig:CharmBroadening}.

For the medium-induced component at first order in opacity, the fractional energy loss in the DGLV formalism was derived in Ref.~\cite{Gyulassy:2000er} and is summarized in Sec.~\ref{sec:infeloss}.  The DGLV prediction \eqref{e:eloss6} indicates a softening of the energy loss relative to the vacuum due to the destructive interferences of the LPM effect: the single logarithm from Eq.~\eqref{e:vacasym} survives, but is further suppressed by a power of the energy in Eq.~\eqref{e:eloss7}.  Moreover, for the energy loss calculation the integrand decays sufficiently fast in the UV that the integral is well behaved even if the integration limits are extended to infinity.  This allows for a straightforward calculation of the high-energy asymptotics which is fairly insensitive to the assumptions made about the integration limits.

As seen in Sec.~\ref{sec:infbroad} and Sec.~\ref{sec:finitebroad}, this simple picture does not extend to the case of transverse momentum broadening, which is much more sensitive to the choice of UV limits due to the weighting by $k_\bot^2$.  The calculation of the radiative (NLO) component of jet broadening in both the DGLV and twist-4 formalisms is predicted to be double-logarithmic, growing as  $\ln^2 \tfrac{E}{\mu}$ as $E \rightarrow \infty$.  However, the \textit{coefficient} of that double logarithm differs substantially between the two formalisms and based on the approximations used to compute it.  The calculation \eq{e:DGLVratio} in the DGLV framework with the assumption of infinite kinematics resulted in a coefficient which was positive.  A direct numerical evaluation of the coefficient obtained from the DGLV framework incorporating the finite kinematic limits obtained a coefficient which was \textit{negative}, as shown in Fig.~\ref{fig:ratioEoMu}.  A comparable calculation using the twist-4 formalism of Refs.~\cite{Kang:2014ela,Xing:2014kpa, Kang:2013raa} in \eq{e:ZBKratio} found a numerical coefficient consistent with \eq{e:DGLVratio}, but with an explicit dependence on the PDFs.  This feedback between the medium-induced branching and the initial hard scattering is an entanglement of initial and final states which is generally neglected in the energy loss approach.

In a detailed tandem analytic-numerical analysis per\-form\-ed in Sec.~\ref{sec:finitebroad} we explored the origin of this discrepancy.  We found that, when including the constraints of finite kinematics, the integration range for the radiative broadening in medium includes \textit{multiple} double-logarithmic regimes whose relative weights depend sensitively on the boundaries of the integration region.  Thus while a naive implementation of the integrals using infinite kinematics predicted a positive coefficient, both the numerical evaluation in Fig.~\ref{fig:ratioEoMu} and the analytic evaluation in \eq{eq:leadingEratio} show that for finite kinematics in DGLV, the coefficient is negative.  

The substantial sensitivity of the jet momentum broadening to the assumptions employed in the calculation raises significant questions about the most theoretically sound basis to study such effects.  For instance, the appearance of a convolution over the PDFs in the twist-4 formalism \eqref{e:ZBKratio} indicates a substantial cross-talk between the initial-state phys\-ics and the final-state modification of the jets.  This feature stands in contradistinction to the assumption common in energy-loss frameworks that the initial- and final-state physics can be factorized.  Moreover, the energy loss kernels are often implemented in the form of a ``medium-modified DGLAP evolution'' applied to the fragmentation functions; this seems difficult to reconcile with the terms in \eqref{e:ZBKratio} which enter as corrections to the initial state PDF.  

On the other hand, there are important questions to ask about the assumptions underlying the collinear twist-4 calculation as well.  For instance, the collinear operators which characterize the double-PDF $T_{q g}$ of \eq{eq:tw4el} are collinear (that is, light-like separated); the Fourier-conjugate momentum variables have been integrated to infinity.  But as we saw in Sec.~\ref{sec:infbroad}, extending the momentum integrals to infinity can drastically change the coefficient of the radiative momentum broadening compared to the full result including finite kinematics.  What then, is the proper role of the finite kinematic limits in the collinear twist-4 calculation, and how can one construct an apples-to-apples comparison of those effects in the two formalisms?  Further, the twist-4 work in principle only includes leading in $\perp/E$ contributions; to capture the full destructive interference of the LPM effect requires a resummation of a subset of higher order in $\perp/E$ terms.  One may also rightly note that, while the calculation of Refs.~\cite{Kang:2014ela,Xing:2014kpa, Kang:2013raa} demonstrated that at NLO the corrections are finite and consistent with an assumption of factorization, no twist-4 factorization theorem has yet been proven.  We note, however, that the presence of additional (potentially nonperturbative) factorization-breaking corrections would only enhance the entanglement of initial and final state, projectile and target.

It's worth now briefly revisiting the discussion of the radiative corrections within the BDMPS-Z approach \cite{Liou:2013qya,Blaizot:2014bha,Iancu:2014kga,Blaizot:2019muz}.  Just like the twist-4 formalism, these works predict a jet broadening from energy loss, as opposed to the jet narrowing we find from the opacity approach.  Recall that in \cite{Liou:2013qya,Blaizot:2014bha,Iancu:2014kga,Blaizot:2019muz}, the high momentum parent parton exists for all time before entering a brick of QCD matter at a finite time.  Therefore these calculations only have positive $\langle p_T^2\rangle(L)$ and do not have any jet broadening in the absence of the brick (in the limit of $\langle p_T^2\rangle(L\rightarrow0) = 0$), unlike in the case of a real hadronic collision.  In even ep and p+p collisions, the hard scattering production process generates a large vacuum shower that broadens the jet; this broadening has been observed experimentally \cite{ZEUS:1997fjy}.  In the opacity expansion approach, that vacuum radiation is destructively interfered with by the stimulated emission of radiation from interactions with the medium, and the result is that the presence of the medium leads to a reduction of the broadening compared to hard scattering in vacuum---an important consequence of the LPM effect.  
What we have shown here is that it's highly non-trivial to determine the arguments of those logarithms for realistic phenomena.  

One may wonder why the three different approaches to jet broadening---twist-4, opacity expansion, and BDMPS-Z---all yield double logarithms.  The answer appears to be that the Sudakov double logarithm is simply ubiquitous; spin-1 radiated quanta generally have a spectrum whose structure goes roughly as $\sim dx/x \, dk_\perp^2/k_\perp^2$.  Unlike in \cite{Liou:2013qya,Blaizot:2014bha,Iancu:2014kga,Blaizot:2019muz} where significant simplifying assumptions lead to a reduction of phase space from 4D (in $x,\,k_T,\,q_T,$ and $\theta_{kq}$) to 2D ($x$ and $k_T$), in Sec. \ref{sec:finitebroad} we found a highly nontrivial competition between \emph{multiple} double-logarithmic regions, in which even the angular integral played a nontrivial role.  A more careful treatment within the BDMPS-Z framework also predicts a jet narrowing \cite{Zakharov:2018rst,Zakharov:2019fov,Zakharov:2020sfx}.  

As noted in \cite{Blaizot:2019muz} a noteworthy feature of some of the work of \cite{Liou:2013qya,Iancu:2014kga} is the use of the size of the QCD medium brick to set an upper limit to the formation time of the emitted radiation.  This cutoff is completely artificial: the radiated quanta could of course form in the vacuum beyond the extent of the brick.  (These derivations are not taken in a finite-sized universe whose extent is given by the length of the QCD medium brick.)  However, in the language of \cite{Liou:2013qya,Blaizot:2014bha,Iancu:2014kga,Blaizot:2019muz}, the Sudakov double logarithm has one logarithmic contribution that scales like $d\tau/\tau$, where $\tau$ is the formation time.  It appears that should those calculations allow their radiated quanta to come on-shell outside of their brick---i.e.\ without the artificial cutoff on the formation time---then their jet broadening prediction would always be infinite.  

The data shown in Fig.~\ref{fig:ALICEdata} on jet momentum broadening in $AA$ collisions from ALICE is ambiguous, being consistent with a broadening, narrowing, or no change relative to the vacuum given current experimental uncertainties.  Clearly \linebreak progress in controlling the uncertainties of the theoretical calculations will be important for discriminating model predictions once more precise data become available.  Outside of heavy-ion collisions, one may also look at jet momentum broadening in cold nuclear matter in $eA$ and $pA$ collisions.  To this end, in Ref.~\cite{Ru:2019qvz} the authors perform a global analysis at leading order in the twist-4 framework (collisional momentum broadening), finding that a nontrivial dependence on $\hat{q}$ with kinematics is required to describe the data.  The data in $eA$ and $pA$ collisions unambiguously show a net broadening compared to vacuum, but strikingly, the HERMES data shown in Fig.~3 of Ref.~\cite{Ru:2019qvz} shows that the amount of broadening \textit{decreases} with increasing current jet energy $\nu$.  We note that this curious energy dependence is qualitatively consistent with the prediction of the opacity expansion for $\Delta \langle p_T^2 \rangle_\mathrm{tot}$ shown in our Fig.~\ref{fig:CharmBroadening}: at low jet energies, the positive contribution of collisional broadening dominates, whereas at high jet energies the growing radiative component reduces the amount of broadening.  Based on these observations, it would be interesting to perform a similar global analysis based on the DGLV / energy loss approach in future work.  

As we have shown here, constructing an apples-to-apples comparison between the energy loss and twist-4 formalisms will require significant progress on the theoretical uncertainties of both theories.  We regard this work as a step toward reconciling the commonalities and differences of the two theoretical frameworks to improve the theoretical description of jet-medium interactions.

\section*{Acknowledgements}

The authors would like to thank Matthias Burkardt, Zhongbo Kang, John Lajoie, Nobuo Sato, Ivan Vitev, and Hongxi Xing for useful discussions.  WAH thanks the South African National Research Foundation and the SA-CERN Collaboration for financial support.  MDS is supported by a start-up grant from New Mexico State University.  HC thanks the Skye Foundation, the Oppenheimer Memorial Trust and the Cambridge Trust for financial support.


\end{document}